\documentclass[twocolumn]{aastex631}

\newcommand{\kms}{\rm km~s^{-1}}
\newcommand{\kmsmpc}{\rm km~s^{-1}~Mpc^{-1}}
\newcommand{\dn}{D_{n}4000}

\usepackage{color}
\usepackage{multirow}
\usepackage{amsmath}
\usepackage{appendix}
\usepackage{hhline}
\usepackage{hyperref}

\shorttitle{HectoMAP sizes}
\shortauthors{Damjanov et al.}

\graphicspath{{./}{figures/}}

\begin{document}

\title{Size and Spectroscopic Evolution of HectoMAP Quiescent Galaxies}

\author{Ivana Damjanov}
\affil{Department of Astronomy and Physics, Saint Mary's University, 923 Robie Street, Halifax, NS B3H 3C3, Canada; \href{mailto:Ivana.Damjanov@smu.ca}{Ivana.Damjanov@smu.ca}}
\affil{Canada Research Chair in Astronomy and Astrophysics, Tier II}
\author{Jubee Sohn}
\affil{Center for Astrophysics | Harvard \& Smithsonian, 60 Garden Street, Cambridge, MA 02138, USA}
\affil{Astronomy Program, Department of Physics and Astronomy, Seoul National University, Gwanak-gu, Seoul 151-742, Republic of Korea}
\author{Margaret J. Geller}
\affil{Center for Astrophysics | Harvard \& Smithsonian, 60 Garden Street, Cambridge, MA 02138, USA}
\author{Yousuke Utsumi}
\affil{SLAC National Accelerator Laboratory, Menlo Park, CA 94025, USA}
\affil{Kavli Institute for Particle Astrophysics and Cosmology, Stanford University, Stanford, CA 94305, USA}
\author{Ian Dell'Antonio}
\affil{Department of Physics, Brown University, Box 1843, Providence, RI 02912, USA}

\begin{abstract}
The HectoMAP survey provides a complete, mass-limited sample of 30,231 quiescent galaxies with $i-$band Hyper Suprime-Cam Subaru Strategic Program (HSC SSP) imaging that spans the redshift range $0.2 <z < 0.6$. We combine half-light radii based on HSC SSP imaging with redshifts and $\dn$ to explore the size -- mass relation, $R_{e} = A \times M_{*}^{\alpha}$, and its evolution for the entire HectoMAP quiescent population and for two subsets of the data. Newcomers with $1.5 < \dn < 1.6$ at each redshift show a steeper increase in $A$ as the universe ages than the population that descends from galaxies that are already quiescent at the survey limit, $z \sim 0.6$ (the resident population). In broad agreement with previous studies, evolution in the size -- mass relation both for the entire HectoMAP sample and for the resident population (but not for the newcomers alone) is consistent with minor merger driven growth. For the resident population, the evolution in the size -- mass relation is independent of the population age at $z \sim 0.6$. The contrast between the sample of newcomers and the resident population provides insight into the role of commonly termed ``progenitor bias" on the evolution of the size -- mass relation.
\end{abstract}


\section{Introduction} \label{sec:intro}

During the last fifteen years, observations have provided an increasingly detailed picture of the evolution of the quiescent galaxy population. Perhaps the most remarkable aspect of the evolution is the size growth at fixed stellar mass for galaxies with masses $\gtrsim 10^{10}{\rm M}_{\odot}$. From a redshift of 1.5 to the present, the growth is a factor of 2.5-4 (\citealp{Trujillo06, Trujillo07, Williams10, Damjanov11, Newman12, Cassata13, Huertas-Company13, vanderWel14, Faisst17, Damjanov19, Mowla19, Mosleh20, Kawin21, Yang21, Barone22, Hamadouche22} and others).

Evaluation of the size growth of the quiescent population is based on the size -- mass relation, a projection of the stellar mass fundamental plane \citep{Hyde09, Bezanson13, Zahid16}. The size -- mass relation for quiescent galaxies with stellar masses $\gtrsim 10^{10}{\rm M}_{\odot}$ is $R_{e} = A \times M_{*}^{\alpha}$ where $R_{e}$ is the half-light radius. Remarkably, the slope, $\alpha$ varies very little with redshift even though $A$ increases substantially (e.g. \citealp{Faisst17, Hamadouche22}). Many evolutionary studies have focused on $z \gtrsim 1$ where the size evolution appears to be more rapid than it is at more recent epochs (e.g., \citealp{Cassata11, Newman12, Damjanov19}). One limitation of these investigations is lack of sufficiently large quiescent samples with complete spectroscopy in the redshift range $0.2 < z < 0.6$. 

The consensus view is that minor mergers are the main driver of quiescent galaxy size evolution at $0.2 < z < 0.6$ and possibly over a larger redshift range (e.g., \citealp{Nipoti09, Naab09, Nipoti12, Newman12, Faisst17, Hamadouche22}). One issue that clouds the interpretation is the necessary distinction between objects that are just becoming quiescent at any particular epoch and those that are residents evolving from earlier epochs. This issue is generally called progenitor bias \citep{vanDokkum01, Carollo13, Belli15, Fagioli16}. For greater specificity in our discussion we distinguish between newcomers (the sources of progenitor bias) and the resident population. 

We explore insights into the size evolution based on the quiescent population in the HectoMAP survey. HectoMAP is a dense red-selected redshift survey covering 55 square degrees of the northern sky \citep{Geller15, Sohn21, Sohn22}. The redshift survey includes a complete mass-limited sample of $\sim 42,500$ quiescent objects at $0.2 < z < 0.6$ selected on the basis of the spectroscopic indicator $\dn$ (Section \ref{spec}). We derive half-light radii for 30,231 of these objects from Subaru Hyper Suprime-Cam (HSC) Subaru Strategic Program (SSP) $i-$band imaging \citep{Aihara22} that covers 80\% of the HectoMAP region (Section \ref{HSC}). The MMT Hectospec spectroscopy enables separation of likely newcomers from the resident population already in place at the limiting redshift, $z = 0.6$. We can thus evaluate the impact of newcomers on the size evolution of the population.

Section \ref{data} describes the data that underlie our analysis. We review the red-selected HectoMAP survey (Section \ref{spec}) and we use the magnitude limited SHELS survey to comparable depth to evaluate the completeness of the HectoMAP quiescent sample (Section \ref{pop}). We review the HSC/SSP photometry and describe our size measurements (Section \ref{HSC}). We use $r-$ and $i-$band Subaru photometry to demonstrate the insensitivity of the size evolution results to the use of $i-$band photometry throughout the redshift range. 

Section \ref{SM} highlights the HectoMAP quiescent galaxy size -- mass relation for the entire complete stellar mass limited sample of $30,231$ quiescent galaxies with publicly available $i-$band HSC/SSP imaging. Because of the large size of the total quiescent sample we can follow the evolution of newcomers (objects with $1.5 < \dn < 1.6$) as the lookback time increases by only $\sim 1.5$ Gyr (Section \ref{new}). We follow the quiescent population that ages from a redshift of $z = 0.6$, the resident population (Section \ref{reside}). We compare the size -- mass relations for this population with the relation for the newcomers and for the full quiescent sample. Finally we assess the agreement between the evolution of these size -- mass relations and broadly favored minor merger models (Section \ref{merge}). We discuss the results in a broader forward looking context in Section~\ref{limit}. We conclude in Section~\ref{con}.

We adopt the Planck cosmological parameters \citep{Planck16} with $H_{0} = 67.74~\kmsmpc$, $\Omega_{m} = 0.3089$, $\Omega_{\Lambda} = 0.6911$ and the AB magnitude system.

\section{The Data} \label{data}

We use the red-selected HectoMAP redshift survey \citep{Geller15, Sohn21, Sohn22} along with HSC/SSP imaging \citep{Miyazaki18, Aihara22} of the HectoMAP region to trace the size evolution of the quiescent population from $z \sim 0.2$ to $z \sim 0.6$. The redshift survey provides the spectral indicator $\dn$ and the stellar masses that we use to define a mass limited sample of $30,231$ quiescent objects covering this redshift range. Because HectoMAP is color-selected, we calibrate the mass selected sample with the complete purely magnitude limited SHELS F2 survey \citep{Geller14, Geller16}. At $z \lesssim 0.2$ the color selection of HectoMAP removes a significant portion of the quiescent population. We thus restrict our analysis to the redshift range $0.2 < z < 0.6$.

\subsection{HectoMAP Spectroscopy}\label{spec}

HectoMAP is a dense, red-selected redshift survey carried out with the Hectospec 300-fiber wide-field instrument on the 6.5-meter MMT \citep{Fabricant05}. The survey covers a total area of 55 deg$^{2}$ in a 1.5 deg wide strip at high declination. The right ascension and declination limits of the survey are: $200 <$ R.A. (deg) $< 250$ and $42.5 <$ Decl. (deg) $< 44.0$ \citep{Sohn21, Sohn22}.

Observations for HectoMAP span more than a decade. We based photometric selection on successive SDSS data releases and we now calibrate the completeness to SDSS DR16 \citep{Ahumada20}. HectoMAP includes a bright survey with $r_{petro, 0} < 20.5$ and a red color selection $(g-r)_{model, 0} > 1$ and a faint survey with $20.5 < r_{petro, 0} \leq 21.3$ and red selection $(g-r)_{model, 0} > 1$ and $(r-i)_{model, 0} > 0.5$. Here $r_{petro, 0}$ refers to the Sloan Digital Sky Survey (SDSS) Petrosian magnitude corrected for Galactic extinction and the color selection is based on the extinction-corrected SDSS model magnitudes. For the faint portion, the $(r-i)_{model, 0}$ cut removes low redshift objects. The cut was necessary because the telescope time required without the cut was prohibitive. We impose an additional surface brightness limit by requiring $r_{fiber, 0} < 22.0$ throughout the HectoMAP survey because red objects below this limit yield very low signal-to-noise spectra. The limiting $r_{fiber,0}$ magnitude corresponds to the extinction-corrected flux within the Hectospec fiber aperture (with $0\farcs75$ radius). 

\citet{Sohn21} and \citet{Sohn22} describe the two data releases that cover the entire HectoMAP redshift survey. These two papers included detailed descriptions and analyses of the survey completeness and uniformity. They also include detailed discussion of the redshifts, the spectral indicator $\dn$, and the stellar mass. Here we very briefly summarize these data along with any issues that might affect the investigation we carry out here.

HectoMAP is a dense redshift survey. There are $\sim 2000$ redshifts per square degree. The median depth of the survey is $z = 0.31$. Within the color selection the survey is 80\% uniformly complete to the limit $r_{petro, 0} = 20.5$. In the faint sample with $20.5 < r_{petro, 0} \leq 21.3$, the sample is 65\% complete and the completeness is somewhat lower near the edges of the survey. The total number of redshifts is 100,097. Table~\ref{tab1} summarizes these properties of the survey\footnote{Note that the numbers in this table are related to the portion of the survey that covers redshift interval $0.2 < z < 0.6$. The total number of spectra in the survey is thus larger.}. 

The typical error in a HectoMAP redshift is $38~\kms$. There is a small offset of $39~\kms$ between Hectospec redshifts and the SDSS \citep{Sohn21}. This offset is irrelevant for the analysis we pursue here. 

We compute stellar masses of HectoMAP galaxies based on SDSS DR16 $ugriz$ photometry. We use the Le Phare software package \citep{Arnouts99, Ilbert06} that incorporates stellar population synthesis models of \citet{Bruzual03}, a \citet{Chabrier03} IMF and a \citet{Calzetti00} extinction law. We explore two metallicities (0.4 and 1) and a set of exponentially decreasing star formation rates. We derive the mass-to-light ratio from the best model and we then use this ratio to convert the observed luminosity into a stellar mass.

\subsection{Quiescent Population Definition and Selection}\label{pop}

As in our previous investigations of the quiescent population, we use the spectral indicator $\dn$ to segregate the quiescent population. The quantity $\dn$ is a flux ratio (in $f_{\nu}$ units) in two bands: $4000–4100$~\AA\ and $3850–3950$~\AA\ \citep{Balogh99}. This ratio is a measure of the age of a stellar population. The index is smaller for objects with a younger population \citep{Kauffmann03}.

For the HectoMAP spectra, the median signal-to-noise around 4000~\AA\ is 4.5. More than 95\% of the HectoMAP spectra have adequate signal-to-noise for robust determination of the index.  Based on repeat measurements, \citet{Geller14} report a typical error of 4.5\% in $\dn$ for the comparable SHELS survey data. \citet{Damjanov22} show (their Figure 3) that in the SHELS survey the typical error in $\dn$ increases with redshift. In the Appendix~\ref{appendix1} we demonstrate that the error in $\dn$ has a negligible effect on analysis of the size -- mass relation and its evolution.

Following previous studies (e.g., \citealp{Vergani08, Damjanov18, Damjanov19, Hamadouche22}) we identify the quiescent population as objects with $\dn > 1.5$. \citet{Damjanov18} use the hCOSMOS redshift survey to test this spectroscopic selection against rest-frame UVJ color selection based on SED fitting to 30 photometric bands covering the range 0.15 - 254 $\mu$m. They find that the spectroscopic and photometric approaches agree well. Furthermore, in the COSMOS field, \citet{Damjanov18} examine the relationship between Zurich Estimator of Structural Types (ZEST; \citealp{Scarlata07}) morphologies and the spectroscopic indicator of galaxy quiescence. About ninety percent of galaxies with $\dn > 1.6$ have elliptical or bulge-dominated morphologies as do $\sim 70 \%$ of the objects with $1.5 < \dn < 1.6$. 

\begin{figure*}[!h]
\centering
\includegraphics[width=0.9\textwidth]{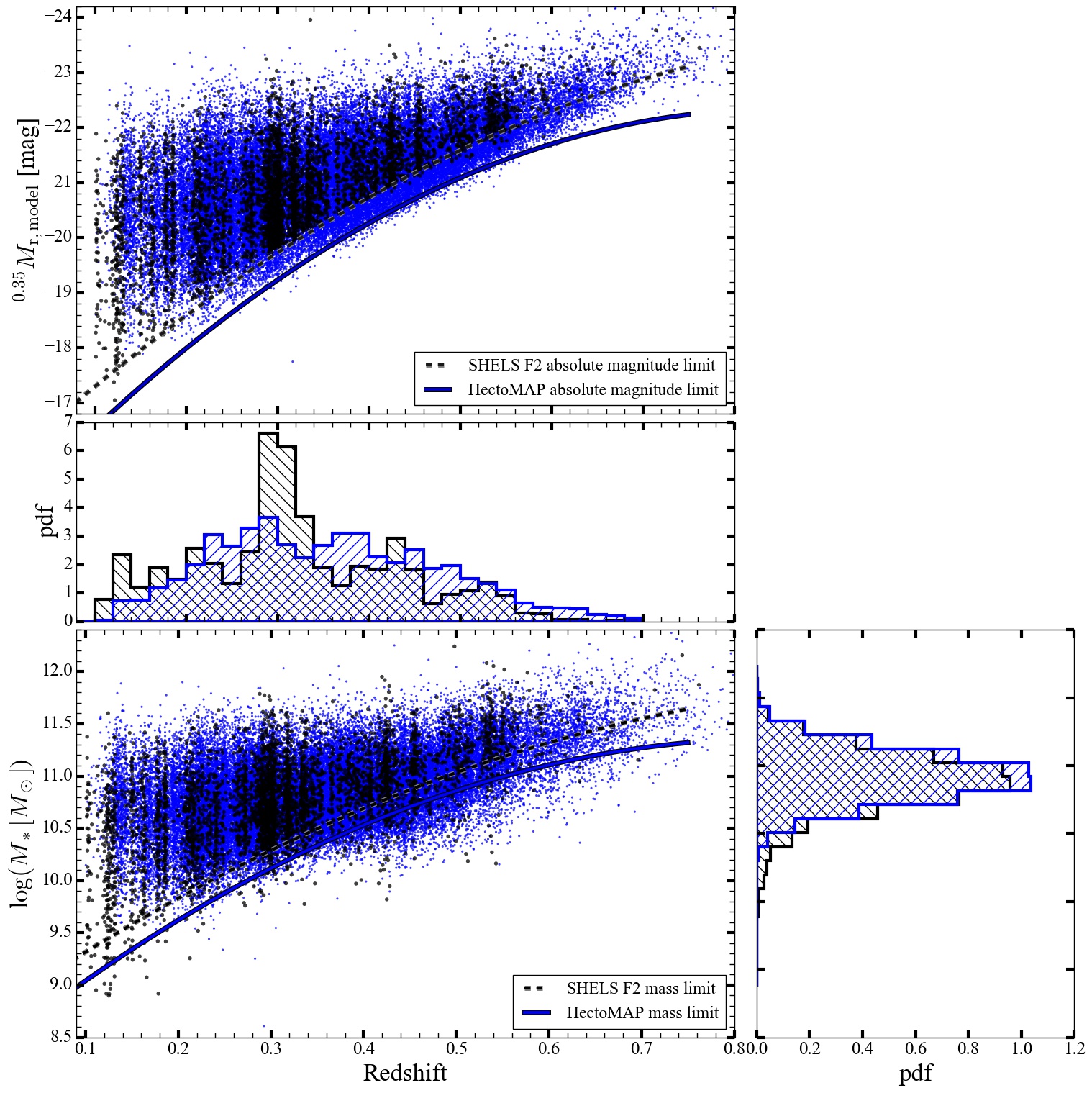}
\caption{Absolute K-corrected $r-$band magnitude ({\it top} panel) and stellar mass ({\it bottom} panel) versus redshift for the quiescent galaxy population in SHELS F2 ($r < 20.75$; black points and black dashed lines) and HectoMAP ($r < 21.3$; blue points and blue dashed lines). The absolute magnitude limits (black dashed and blue solid lines in the top panel), corresponding to the observed K-corrected $r-$band absolute magnitude limits. The lower panel shows the SHELS F2 and HectoMAP stellar mass limits (black dashed and blue solid lines in the lower panel) derived by incorporating the average mass-to-light ratio for SHELS F2 and HectoMAP galaxies at each redshift. The horizontal and vertical histograms show probability density distributions for the stellar mass and redshift of SHELS F2 (black) and HectoMAP (blue) galaxies.
\label{f1}}
\end{figure*}

Construction of a mass complete sample of quiescent galaxies is a crucial foundation for exploring the evolution of the population. We use the SHELS F2 sample \citep{Geller14, Geller16} to calibrate HectoMAP. SHELS F2 is somewhat shallower than HectoMAP, but it is not color-selected. In magnitude limited samples like SHELS and HectoMAP the apparent magnitude limit imprints a brighter limiting absolute magnitude along with an increasing limiting stellar mass as the redshift increases. The upper panel of Figure \ref{f1} shows the K-corrected $r-$band absolute magnitudes along with magnitude limits for both SHELS (black points and black dashed line) and HectoMAP (blue points and blue dashed curve) quiescent galaxies with $\dn > 1.5$ as a function of redshift. The apparent magnitude limit for SHELS is $r = 20.75$; for HectoMAP the limit is $r = 21.3$. For HectoMAP, the color cuts are responsible for the departures from the blue dashed curve that indicates the limiting absolute K-corrected $r-$band magnitude. The middle panel shows the redshift distributions for the two surveys.

We translate the samples limited in absolute magnitude into samples limited in stellar mass. We make the conversion by computing the average mass-to-light ratio in SHELS and HectoMAP at each redshift. The dashed black (SHELS) and blue (HectoMAP) curves show the resulting limits. Here again the black and blue points show quiescent galaxies in SHELS and HectoMAP, respectively. Some objects lie below the limits as a result of photometric errors. Table~\ref{numbers} shows that the fraction of galaxies below the limit is $\lesssim10\%$ for both SHELS F2 and HectoMAP samples. 

\begin{figure}[!h]
\centering
\includegraphics[width=0.45\textwidth]{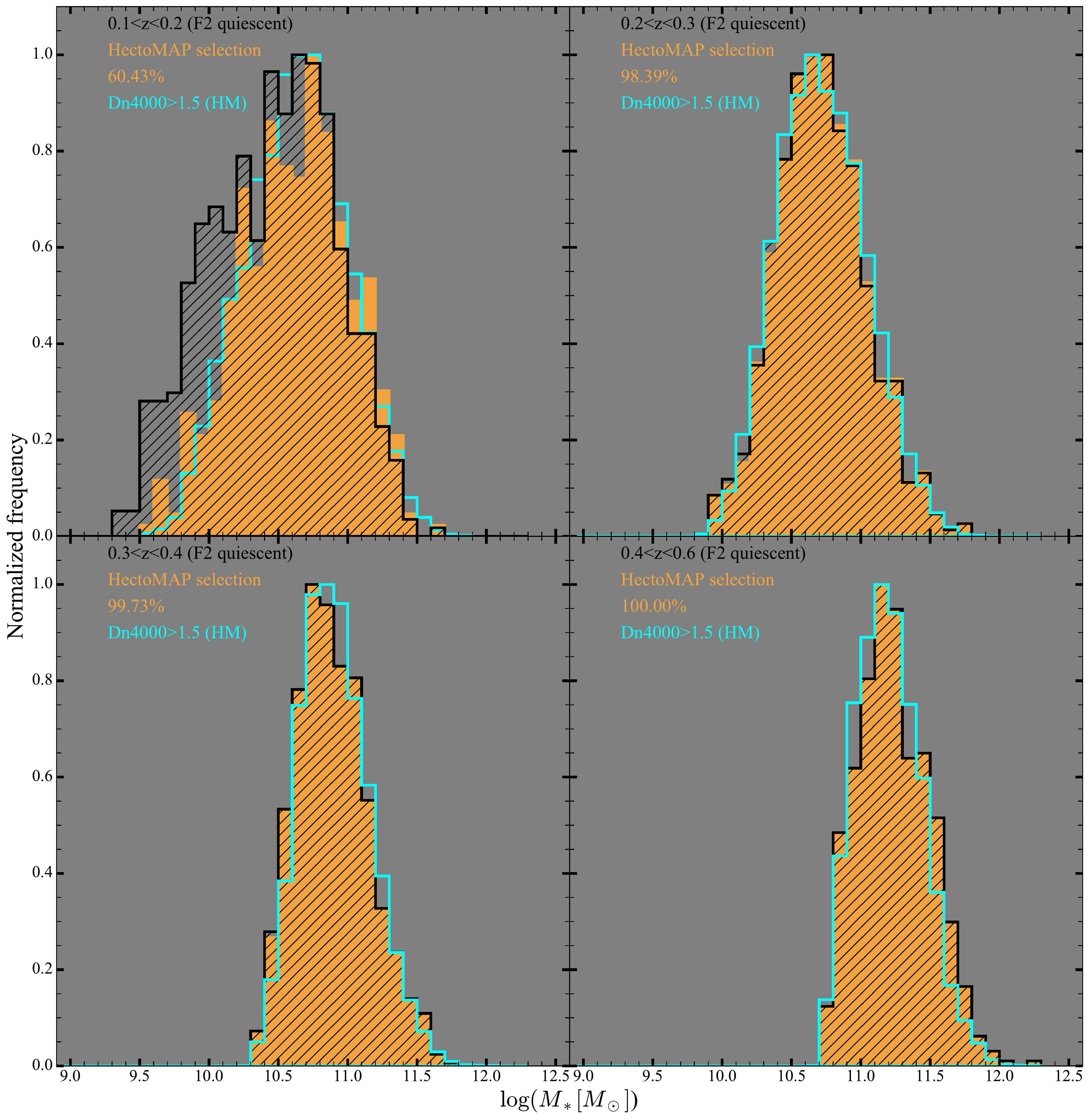}
\caption{Mass distribution of SHELS F2 quiescent galaxies (black histogram), color-selected SHELS F2 quiescent galaxies (orange histogram, $(g-r)>1$ for $r<20.5$ and $(g-r)>1, (r-i)>0.5)$ for $20.5<r<20.75$), and HectoMAP quiescent galaxies (cyan histogram) for four redshift bins covering the $0.1<z<0.6$. Galaxy samples are mass complete above the completeness limit of the SHELS F2 survey (dashed black line in the bottom panel of Figure \ref{f1}). For $0.2<z<0.6$, the red-selection of the HectoMAP survey includes $>98\%$ of the quiescent galaxies in the purely magnitude limited SHELS F2 survey.
\label{f2}}
\end{figure}

The side panel in Figure \ref{f1} shows the aggregate distributions of stellar masses for the SHELS and HectoMAP samples normalized to unit area for each histogram. The mass limited samples of SHELS and HectoMAP quiescents contain, respectively, 4217 and 52593 objects. The distributions are indistinguishable. We examine these distributions further in Figure \ref{f2}.

\begin{deluxetable*}{lcc}\label{tab1}
\tabletypesize{\small}
\tablecaption{Number of galaxies in the SHELS F2 and HectoMAP samples}
\label{numbers}
\tablewidth{7in}
\tablehead{
\colhead{Sample} & \multicolumn{2}{c}{Number of Galaxies}\\
\colhead{} & \colhead{F2} & \colhead{HectoMAP} }
\colnumbers
\startdata 
{SDSS $r-$band limit [mag]} & 20.75 & 21.3 \\
Color selection & None & \multirow{2}{*}{}$(g-r)>1$ for $r<20.5$\\ && $(g-r)>1 \& (r-i)>0.5$ for $r>20.5$\\
$z$ & (0.1, 0.6) & (0.2, 0.6) \\
$N_\mathrm{spec}$ & 10848 &  64240\\
$N_{\mathrm{spec},\, \mathrm{D}_n4000}$ & 10794\tablenotemark{a} & 63317\tablenotemark{a} \\
$N_{\mathrm{D}_n4000>1.5}$ & 4622 & 46367 \\
$N_{\mathrm{D}_n4000>1.5,\, M_\ast}$ & 4602 & 46152\\
$N_\mathrm{mass\, limited\, sample}$ & 4088 & 42496\\
$N_{\mathrm{mass\, limited\, sample}, R_e}$ & 4595\tablenotemark{b} & 30231\tablenotemark{c} \\
\enddata
\tablenotetext{a}{Criteria for reliable $\dn$ measurements are: 1) $\dn > 0$, 2) $\dn < 3$, 3) $\Delta(\mathrm{D}_n4000)<0.2\times\mathrm{D}_n4000$}
\tablenotetext{b}{The $4$~deg$^2$ area of the SHELS F2 survey is fully covered by the HSC $i-$band imaging \citep{Damjanov19}.}
\tablenotetext{c}{The HSC/SSP imaging of the HectoMAP in $i-$band covers $\lesssim 80\%$ of the spectroscopic survey area.}
\end{deluxetable*}

In contrast with the \citet{Damjanov19} analysis of the complete SHELS survey, the red selection of HectoMAP may affect the completeness of mass limited samples of quiescent HectoMAP galaxies.

Figure \ref{f2} explores the completeness of the HectoMAP quiescent population. We use the SHELS survey as a basis for assessing the completeness of HectoMAP. For SHELS galaxies with $r < 20.75$ we first show the quiescent subsample without color selection (black histogram). We then apply the HectoMAP color selection (orange histogram). Figure \ref{f2} shows the number of quiescent galaxies as a function of stellar mass in successive redshift intervals. Within the lowest redshift interval (upper left panel), the total sample (black histogram) includes a tail toward lower stellar mass that is absent in the red-selected sample. These objects are younger (and bluer) than those included in the red-selected sample. This effect is also visible in Figure \ref{f1}.

In the three redshift intervals spanning the redshift range $0.2 < z < 0.6$, the red-selected SHELS samples (orange histograms) are essentially indistinguishable from the total (black histograms); they contain 98-100\% of the galaxies in the total sample. In this higher redshift range, the objects with low stellar mass that populate the tail of the black histograms in the upper left panel are fainter than the limiting magnitude of the SHELS survey. The cyan histograms in Figure \ref{f2} show distributions for HectoMAP quiescents in the same redshift intervals we explore for SHELS. The agreement is excellent for $z \gtrsim 0.2$. Thus HectoMAP provides an essentially complete mass limited quiescent sample over this range. This large sample of $42,496$ quiescent systems provides a powerful platform for examining the evolution of this population. 

The $41$~deg$^{2}$ region of HectoMAP we explore here is large enough that the resulting number densities of quiescent galaxies should be insensitive to the impact of cosmic variance. The expected impact of cosmic variance is only 3.5\% \citep{Driver10,Geller16}.

\begin{figure*}
\begin{centering}
\includegraphics[width=0.9\textwidth]{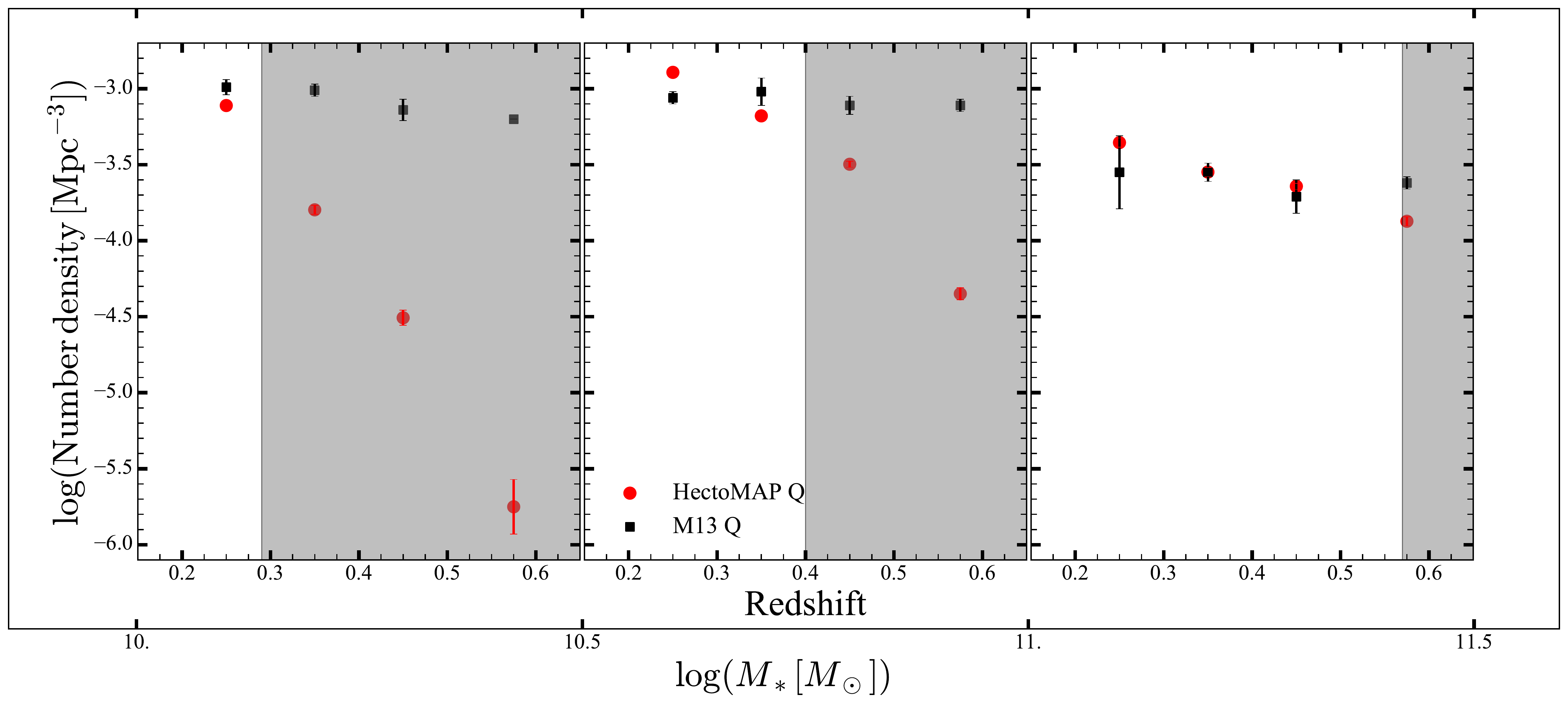}
\caption{Number density of quiescent galaxies in HectoMAP (red symbols) as a function of redshift in bins of stellar mass compared with the number density evolution of galaxies in the PRIMUS survey \citep[][black simbols]{Moustakas13}. In the white area the quiescent HectoMAP sample is $> 99\%$ complete at the lower stellar mass limit of the bin. 
\label{f3}}
\end{centering}
\end{figure*}

Figure \ref{f3} compares the number density of HectoMAP quiescents (red symbols) with the number densities from the 5.5 deg$^2$ PRIMUS grism survey (black points, \citealp{Moustakas13}). PRIMUS quiescents are derived from the star formation rate versus stellar mass diagram, but \citet{Damjanov19} show that this difference in selection does not affect the relative number densities for PRIMUS relative to the complete SHELS F2 redshift survey $\dn$ selection. For the PRIMUS survey the expected impact of cosmic variance is 5.5\%.

For the redshift range $0.2< z < 0.6$ we compare the number densities for three mass ranges. At $z < 0.2$ the red color selection of HectoMAP biases the density estimate because bluer quiescent objects are missing from the sample as noted earlier. In the unshaded regions of Figure \ref{f3} where the HectoMAP samples are mass complete, the number densities of the two surveys agree within a factor of $\lesssim 1.5$. In the highest mass bin (right panel), the agreement is excellent over the entire redshift range we examine in the following analysis.

\subsection{Sizes from HSC SSP Imaging} \label{HSC}

We measure sizes for HectoMAP galaxies with redshifts following the procedure described in detail in \citet{Damjanov19} for the SHELS F2 field of the SHELS survey \citep{Geller14, Geller16}. The $r-$ and $i-$band images for the HectoMAP region are part of the HSC/SSP survey \citep{Miyazaki18}. The HSC $i-$band images of the Third Public Data Release \citep[PDR3,][]{Aihara22}, with median seeing of $\sim 0.55$ arcsec cover 41~deg$^{2}$ ($\lesssim 80\%$ of the survey area) of the HectoMAP region. 

The $r-$band images cover the full HectoMAP region, but the typical seeing is $\sim 0.2^{\prime\prime}$ worse than the $i-$ band seeing. The $r-$band seeing is also less uniform\footnote{https://hsc-release.mtk.nao.ac.jp/doc/index.php/quality-assurance-plots}. We base our analysis on the better and more uniform $i-$band seeing. The results are then directly comparable with the analysis of $i-$band sizes for the SHELS F2 field \citep{Damjanov19}.

The image processing is based on the hscPipe system \citep{Bosch18}, the standard pipeline for the HSC/SSP. The pipeline facilitates reduction of individual chips, mosaicking, and image stacking. The HSC/SSP DR3 offers two sets of co-added images that are based on different approaches to sky subtraction \citep[local and global, see Figure~8 of][]{Aihara22}. Here we select to use $i-$band coadds with global background subtraction where extended wings of bright objects are better preserved.   

As in our earlier exploration of quiescent galaxies size and spectroscopic evolution \citet{Damjanov19}, we estimate the sizes of HectoMAP galaxies based on single S\'ersic profile models \citep{Sersic68}. \citet{Damjanov19} describe the approach and compare the results with previous size measurements in detail; we briefly summarize the procedure here.

\begin{figure}[h]
\begin{centering}
\includegraphics[width=0.45\textwidth]{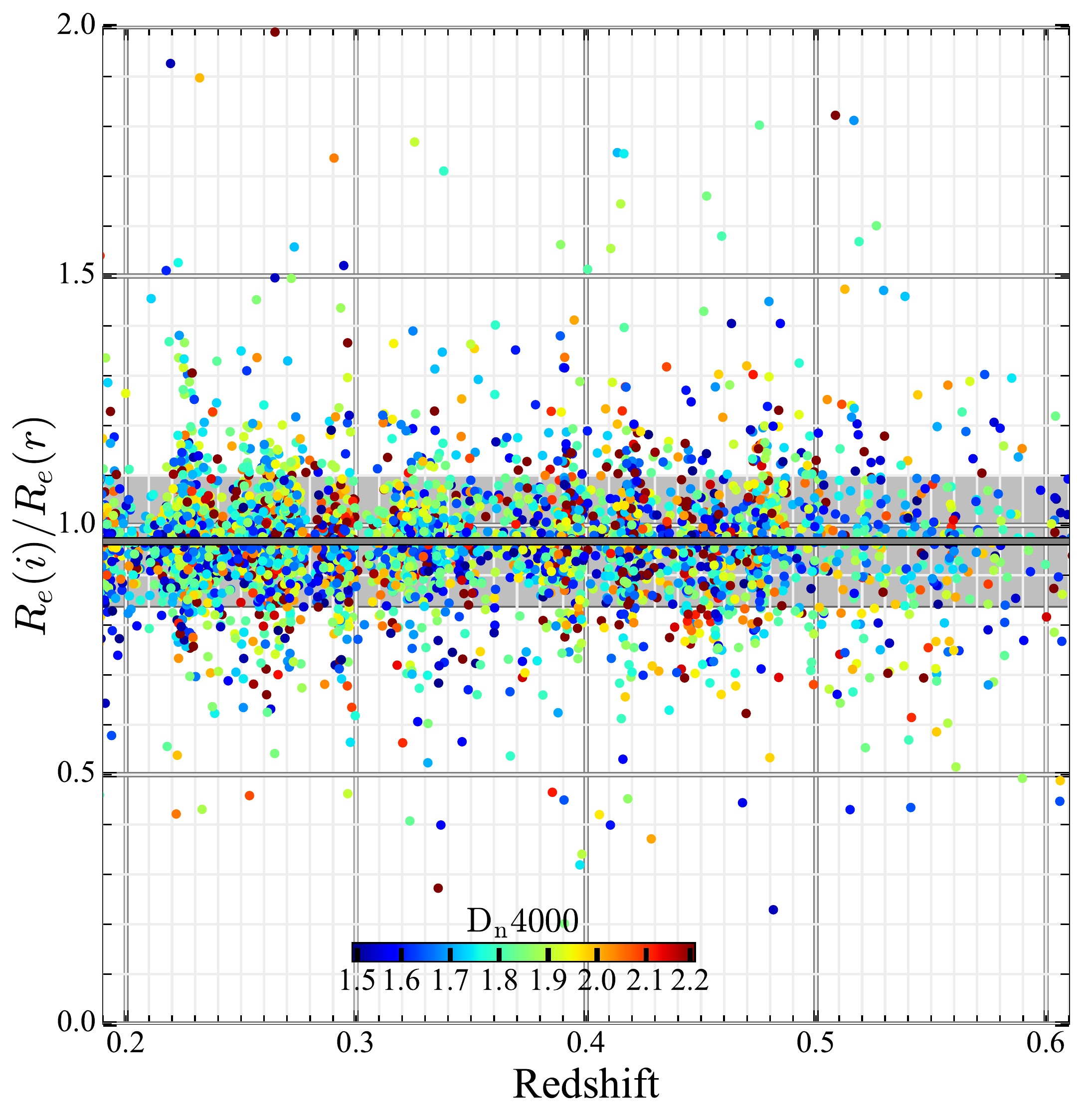}
\caption{The ratio of $i-$band to $r$-band sizes along the galaxy major axis as a function of redshift. The subsample show here includes 4699 HectoMAP galaxies ($\sim10\%$ in the quiescent sample from Table~\ref{tab1}). The symbol for each galaxy is color-coded by the $\dn$ index (proxy for the average stellar population age). The average ratio of two sizes is $0.97$ (dark grey line) and the scatter (the standard deviation of the ratio distribution) is $\pm 0.13$ (grey band). This scatter is essentially the sum in quadrature of the errors in the measurements in each band ($0.07$ for $i$ and $0.11$ for $r$). The (small) difference in sizes is not a function of either redshift or galaxy average stellar population age.  
\label{f4}}
\end{centering}
\end{figure}

We use the SExtractor software \citep{Bertin96} to measure the photometric parameters of galaxies with redshifts in HectoMAP including the galaxy half-light radius, ellipticity, and S\'ersic index. These parameters are based on two-dimensional (2D) modeling of the galaxy surface brightness profile following a three-step process: (a) in its first run, SExtractor provides a catalog of sources that includes star–galaxy separation; (b) the PSFex software \citep{Bertin11} combines clean sets of point sources from the initial SExtractor catalog to construct a set of spatially varying point-spread functions (PSFs) that are the input parameters for (c) the second SExtractor run that provides a catalog with morphological parameters for all detected sources.

Within the \texttt{SPHEROID} model, the SExtractor fitting procedure provides S\'ersic profile parameters and the best-fit model axial ratio b/a (SExtractor model parameter \texttt{SPHEROID\_ASPECT\_WORLD}). The angular diameter distance implied by the spectroscopic redshift of each galaxy translates the galaxy angular size $\Theta_e$ (SExtractor model parameter \texttt{SPHEROID\_REFF\_WORLD}) into the half-light radius along the galaxy major axis  $R_e$ in kpc (e.g., \citet{Hogg99}, and references therein). 

We apply our size measurement procedure to the $i-$band imaging. We also analyze $\sim9$~deg$^2$ of the HSC SSP $r-$band imaging to test for sensitivity to the use of the $i-$band data throughout the redshift range $0.2 < z < 0.6$. 

Figure \ref{f4} shows the ratio $R_{e}(i)/R_{e}(r)$ for 4699 HectoMAP quiescent galaxies covering the entire redshift range we analyze. The $i-$band sizes are slightly smaller than the $r-$band sizes as expected. The solid line shows that the mean ratio, 0.97, does not vary measurably with redshift. 

The gray band shows the 1$\sigma$ error in the ratio, $\pm 0.13$, essentially the sum in quadrature of the total errors in the individual $i-$ and $r-$band size measurements. In each band the total error ($0.07$ for $i$ and $0.11$ for $r$) is a combination of internal errors for individual measurements and the average external error based on galaxies present in multiple HSC SSP cutouts \citep[see][]{Damjanov18}. 

We use the spectral indicator $\dn$ to select the quiescent population and to trace its evolution. Figure \ref{f4} shows the $\dn$ for each galaxy in the $i-$ to $r-$ band size comparison. There is no evident dependence of the size ratio on $\dn$. 

The small difference between the $i-$band and $r-$band sizes and the lack of systematics as a function of either redshift or $\dn$ enable use of the full HectoMAP region coverage in the $i-$band as a robust platform for exploring the size evolution of the quiescent population. 

\section{The size -- mass Relation}\label{SM}

The size -- mass relation is a fundamental ingredient of the evolutionary picture for the quiescent population. Previous investigators demonstrate that the slope of the relation is essentially redshift independent for $z \lesssim 1.5$ \citep[e.g,][]{Newman12,vanderWel14,Faisst17,Mowla19,Barone22}. As the universe ages, the normalization of the relation increases reflecting the size growth of the population \citep[][and others]{Daddi05,Trujillo06,Trujillo07,vanDokkum10,Damjanov11,Newman12,Huertas-Company13,vanderWel14,Huertas-Company15,Faisst17,Damjanov19,Mowla19,Kawin21,Hamadouche22}.

\citet{Zahid17} demonstrate that the size -- mass relation is a function of $\dn$. They use the SDSS to demonstrate that older objects with large $\dn$ have smaller sizes. Quiescent galaxies with successively smaller $\dn$ trace parallel size -- mass relations that shift toward larger sizes for smaller values of $\dn$. \citet{Damjanov19} use SHELS F2 to show that at redshifts $0.2 \lesssim z\lesssim 0.5$ there is also an anti-correlation between $\dn$ and size.

\begin{figure*}
\begin{centering}
\includegraphics[width=0.9\textwidth]{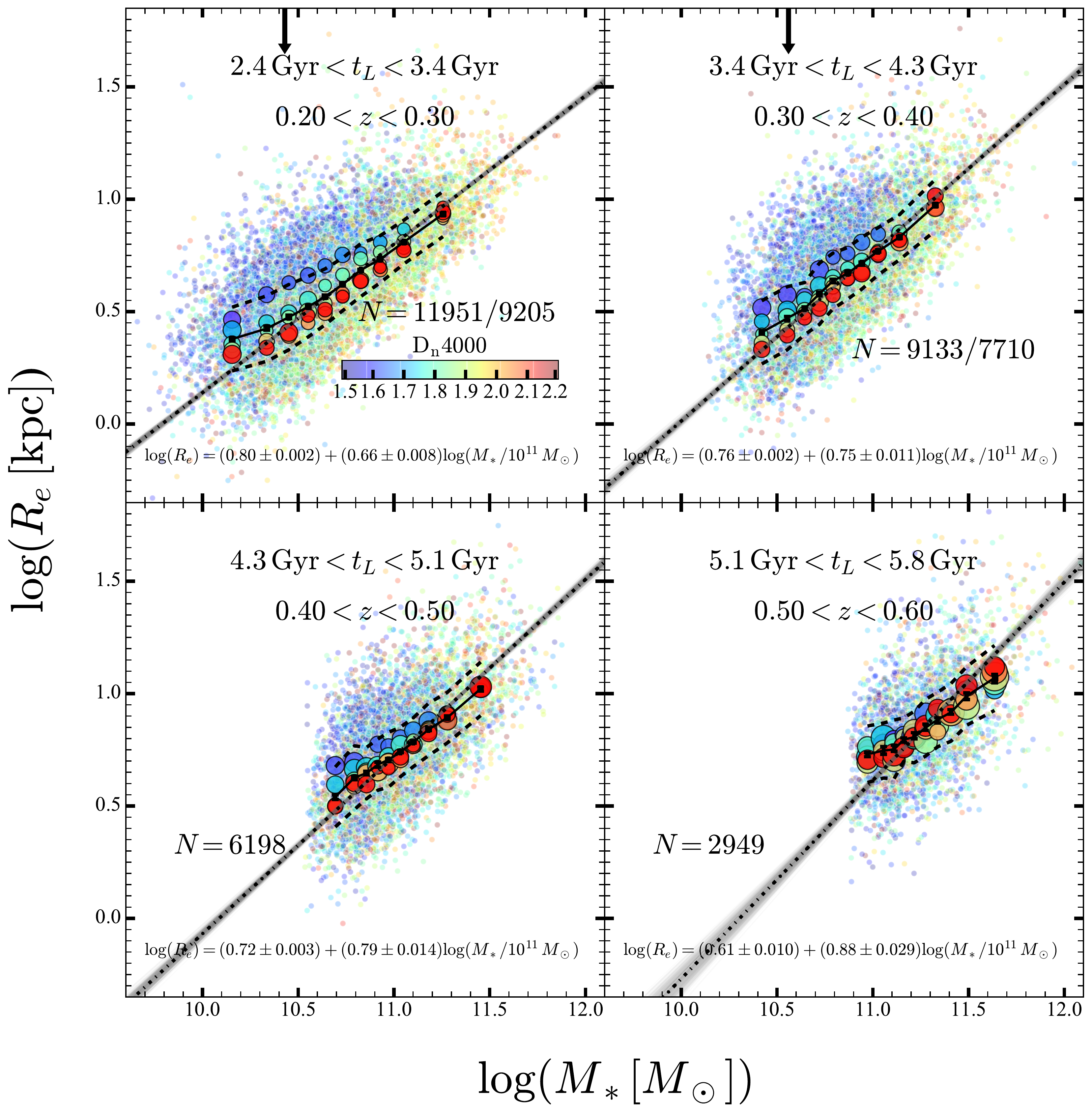}
\caption{Median circularized half-light radius as a function of stellar mass and $\dn$ for quiescent HectoMAP galaxies in four redshift/lookback time intervals. The centers of the large circles indicate the median mass and size of galaxies in 2D bins of stellar mass and $\dn$. The circle color indicates the median $\dn$ in the bin. The size of each circle is proportional to the typical bootstrapped error in the median size. In each panel the solid black line connects the median sizes in 10 equally populated mass bins. The dashed lines show the median absolute deviation. The central black dash-dotted line and gray region traces the best-fit linear relation between stellar mass and size and the associated uncertainty. The background colored points represent individual quiescent HectoMAP galaxies in the mass limited sample appropriate for each panel. The two black arrows indicate the stellar mass where the slope of size -- mass relation changes. We fit the size-stellar mass relation for objects above this pivot point in each redshift bin with an arrow. The pivot mass appears to evolve with redshift. In the higher redshift bins ($z > 0.4$) our sample is above the stellar mass region of the pivot point.
\label{f5}}
\end{centering}
\end{figure*}

Here we explore the constraints on the size -- mass relation provided by the larger HectoMAP quiescent sample. Figure \ref{f5} shows the size -- mass relation for HectoMAP quiescent objects. The panels show the size -- mass relations for mass limited samples. In each panel the black line connects the median values of the relation.

In the two upper panels we fit the median values only for objects with stellar mass above the pivot point indicated by the black arrow. We determine the pivot point by using the software package \texttt{pwlf} \citep{pwlf} to fit a continuous piecewise linear function limited to two segments. The black dash-dotted line in Figure~\ref{f5} shows the fit given in the legend of each panel. The gray region shows the small uncertainty in the fit. The pivot stellar mass appears to evolve with redshift. In the higher redshift bins ($z > 0.4$) shown in the two lower panels, all of the objects in the samples have stellar masses that exceed the pivot point.

To make a clean comparison between Figure \ref{f5} with \citet{vanderWel14}, we combine all of the galaxies in the mass-complete sample covering the redshift range $0.2 < z < 0.5$ and repeat fitting the size -- mass relation. For this ensemble sample the slope is $0.67 \pm 0.01$, which is fully consistent the results of \citet{vanderWel14}, $0.75 \pm 0.06$. \citet{Mowla19} derive a shallower slope, $0.48 \pm 0.03$, based on HST observations of COSMOS galaxies in the same redshift range, but the error in both space-based studies is larger. In contrast with previous work, our large quiescent sample provides a tighter constraint on the slope. The zero points are also fully consistent: we derive $\log(A)=0.57\pm0.01$ for a quiescent galaxy at $5 \times 10^{10}$ M$_{\odot}$ and \citet{vanderWel14} report $0.60 \pm 0.02$. 

We explore variations in the size -- mass relation with redshift/lookback time further in Section \ref{reside} where we compare these results with other subsamples of the data. In Section \ref{reside} we also examine both the scatter around the fit and the intrinsic scatter in the data as a function of redshift. Again we compare these bulk results with various subsamples of the data.

HectoMAP spectroscopy enables clean segregation of the samples in Figure \ref{f5} based on the spectral indicator $\dn$. In each panel of Figure \ref{f5} the colored dots show the median size as a function of stellar mass color coded by $\dn$. The size of each colored circle indicates the error. Comparison with Figure~9 of \citet{Damjanov19} (based on the order of magnitude smaller SHELS F2 dataset, Table~\ref{tab1}) illustrates the power of large spectroscopic samples for identifying trends in a multi-dimensional parameter space like the stellar mass - size - stellar population age proxy ($D_n4000$) for the quiescent population we explore. 

There are several interesting qualitative features of the distributions of $\dn$-encoded points in Figure~\ref{f5}. As the universe ages, the spread between relations for different $\dn$ values increases. The increase in size with decreasing $\dn$ \citep{Zahid17} becomes more evident. These trends result from (1) the broader stellar mass range sampled at lower redshift and (2) the galaxy population entering the quiescent phase at lower redshift and then evolving. Although the size -- mass relation depends on $\dn$, the size -- mass relations for all of the $\dn$ subsamples lie within the median absolute deviation (MAD) shown by the dashed lines.

The size of the quiescent HectoMAP sample enables extraction of other subsamples that inform the physical interpretation of the size growth of the population. In the following sections we describe the derivation and analysis of subsamples that follow the evolution of the population from the highest redshift we access, $z= 0.6$.

\section{Drivers of the Size Growth of the Quiescent Population}\label{drivers}

Processes that drive the size growth of the quiescent population fall into two broad categories: (1) progenitor bias resulting from larger, younger objects that are newcomers to the quiescent population \citep[e.g.,][]{Newman12,Carollo13,Beifiori14,Belli15,Fagioli16,Charbonnier17,Tacchella17,Suess19,Damjanov19,DiazGarcia19,Hamadouche22} and (2) processes that produce growth of the resident population that is already quiescent \citep[e.g.,][]{Newman12,Lopez-Sanjuan12,Belli15,Faisst17,Zahid19,Suess19, DiazGarcia19,Damjanov22,Hamadouche22}. 

In the following discussion we refer to newcomers (rather than continuing to use the term progenitor bias) and resident quiescent populations to clarify the discussion. We minimize the use of progenitor bias terminology to avoid confusion between newcomers at each redshift and higher redshift precursors. 

We demonstrate that the large sample, extended redshift range, and availability of $\dn$ provides limits on the relative contributions of newcomers and processes that produce growth of the aging residents in the redshift range $0.2 < z < 0.6$. From here we refer to objects with $1.5 < \dn < 1.6$ as newcomers. We track the size -- mass relation for the resident quiescent population that descends from $z = 0.6$ progenitors. We compare the results with the size -- mass relation for the entire quiescent population to measure the impact of objects that join the quiescent population at $z \lesssim 0.5$. The objects that join include newcomers at $z \lesssim 0.5$ and their descendants.

\subsection{Newcomers (Progenitor Bias)} \label{new}

\begin{figure*}
\begin{centering}
\includegraphics[width=0.9\textwidth]{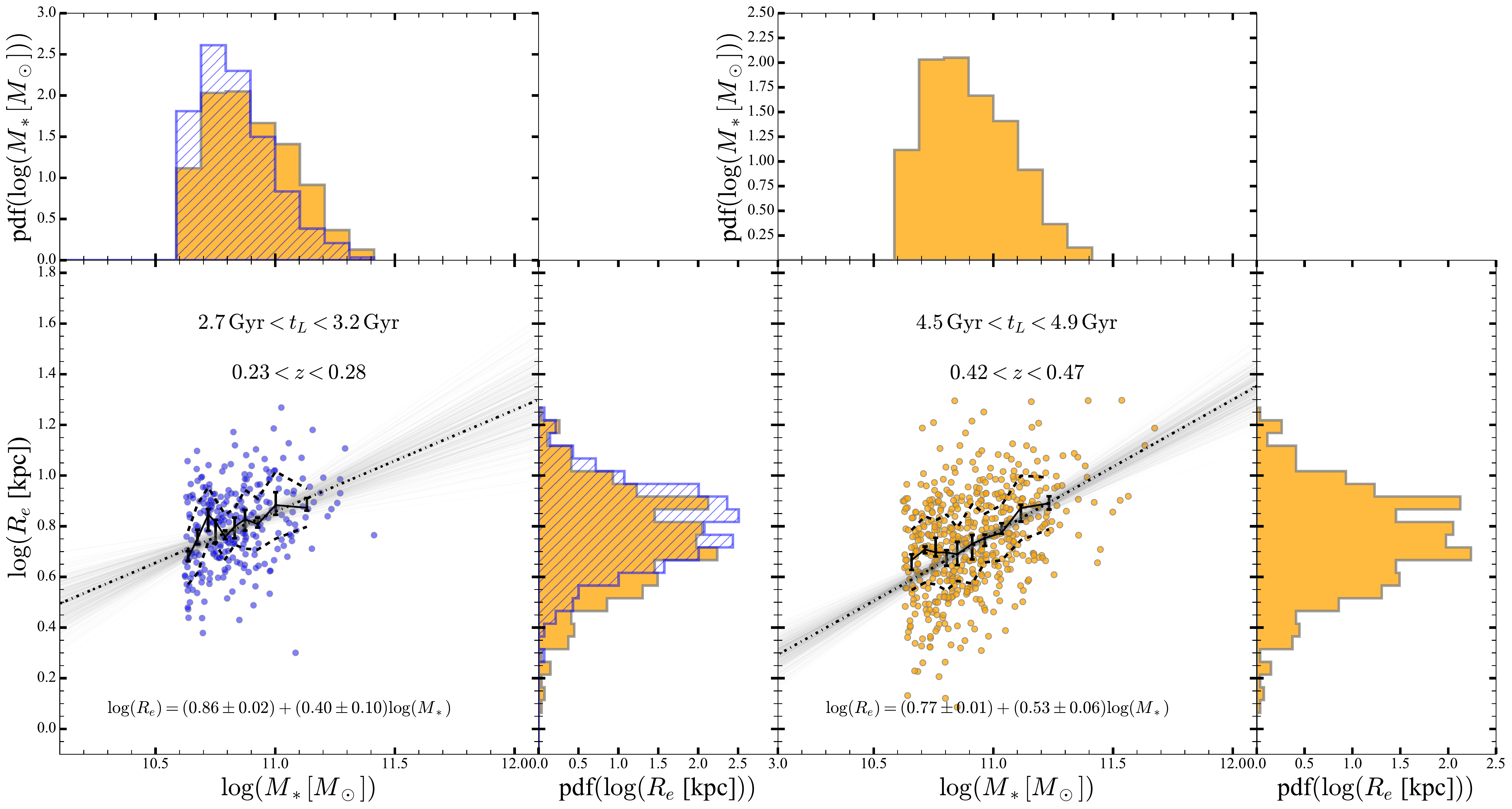}
\caption{Half-light radius along the major axis as a function of stellar mass for newcomers ($1.5 < \dn < 1.6$) HectoMAP galaxies in two narrow redshift bins separated by $\sim1.5$~Gyr. In each panel the solid black line connects the median sizes in equally populated mass bins covering the mass range where the survey is complete at the higher redshift. The errors in the median values are bootstrapped. The central black dash-dotted line and gray region trace the best-fit linear relation between stellar mass and size and the associated uncertainties. The top histograms show normalized stellar mass distributions and the side histograms show normalized size distributions for young quiescents  galaxies in the stellar mass range covered by the higher redshift bin. The top and side histograms associated with the low redshift sample (left; blue histograms) superimpose the mass and size distributions for the higher-redshift sample (orange) in the same mass range. Note the small $R_e<2.5$~kpc in the higher redshift sample.
\label{f6}}
\end{centering}
\end{figure*}

As the universe ages, galaxies that transition from their star-forming phase to quiescence are larger \citep[e.g.,][and many others]{PaulinoAfonso17,Genel18}. To demonstrate the impact of newcomers for $z < 0.6$ we compare the size -- mass relations for newcomers onto the red sequence ($1.5 < \dn < 1.6$) in two narrow, but widely separated redshift bins (Figure \ref{f6}). We select bins within the range covered by HectoMAP because all of the size measurements come from the exquisite HSC imaging. The redshift bins that form the basis for our test are $0.23 < z < 0.28$ and $0.42 < z < 0.47$. These bins bracket an $\sim 1.5$ Gyr change in the age of the universe.

In the range $0.42 < z < 0.47$ the survey is complete for stellar masses at $\log (M_*/M_{\odot}) \gtrsim 10.6$. The black solid line in each scatter plot connects the median size as a function of stellar mass for this size range in both redshift intervals. The bins are equally populated and the errors are bootstrapped. The legend in Figure \ref{f6} gives the two fits. The slopes are the same within the error, but the normalization differs by $\Delta\log(R_{e, 1.5-1.6}) = 0.11 \pm 0.025$. The lower redshift population entering quiescence is larger as expected.

The normalized histograms (probability density functions or pdf) above and to the right of the scatter plots show the stellar mass and $R_e$ distributions, respectively. The histograms that accompany the low redshift panel superimpose the low (blue) and high (orange) redshift pdfs. At the lower redshift there are relatively more objects with lower stellar mass. The $R_e$ pdfs show a striking tail toward small $R_e$ that distinguishes the higher redshift population. This comparison identifies larger recently quenched objects that are joining the quiescent population at later cosmic times as the drivers of the size evolution of the newcomer or progenitor population.

\subsection{Size Growth of an Aging Resident Population}\label{reside}

A range of processes including major mergers, minor mergers, and adiabatic expansion \citep[e.g.,][]{Fan08,Fan10,vanDokkum10,Trujillo11,Lopez-Sanjuan12,Cimatti12,Huertas-Company13} may contribute to the size growth of the quiescent population. For $z \lesssim 1$, several lines of investigation suggest that minor mergers are the primary driver of growth once newcomers are taken into account \citep{Newman12,Beifiori14,Ownsworth14,Belli15,Buitrago17,Zahid19,Hamadouche22}.

We explore the evolution of the resident population by using $\dn$ to connect these resident quiescent galaxies (descendants) at $z < 0.5$ with their progenitors (or ancestors) at $0.5 < z < 0.6$. Once star formation ceases, $\dn$ increases monotonically. We identify $1.5 < \dn < 1.6$ with the cessation of star formation and entry into the quiescent phase.

\begin{figure}
\begin{centering}
\includegraphics[width=0.45\textwidth]{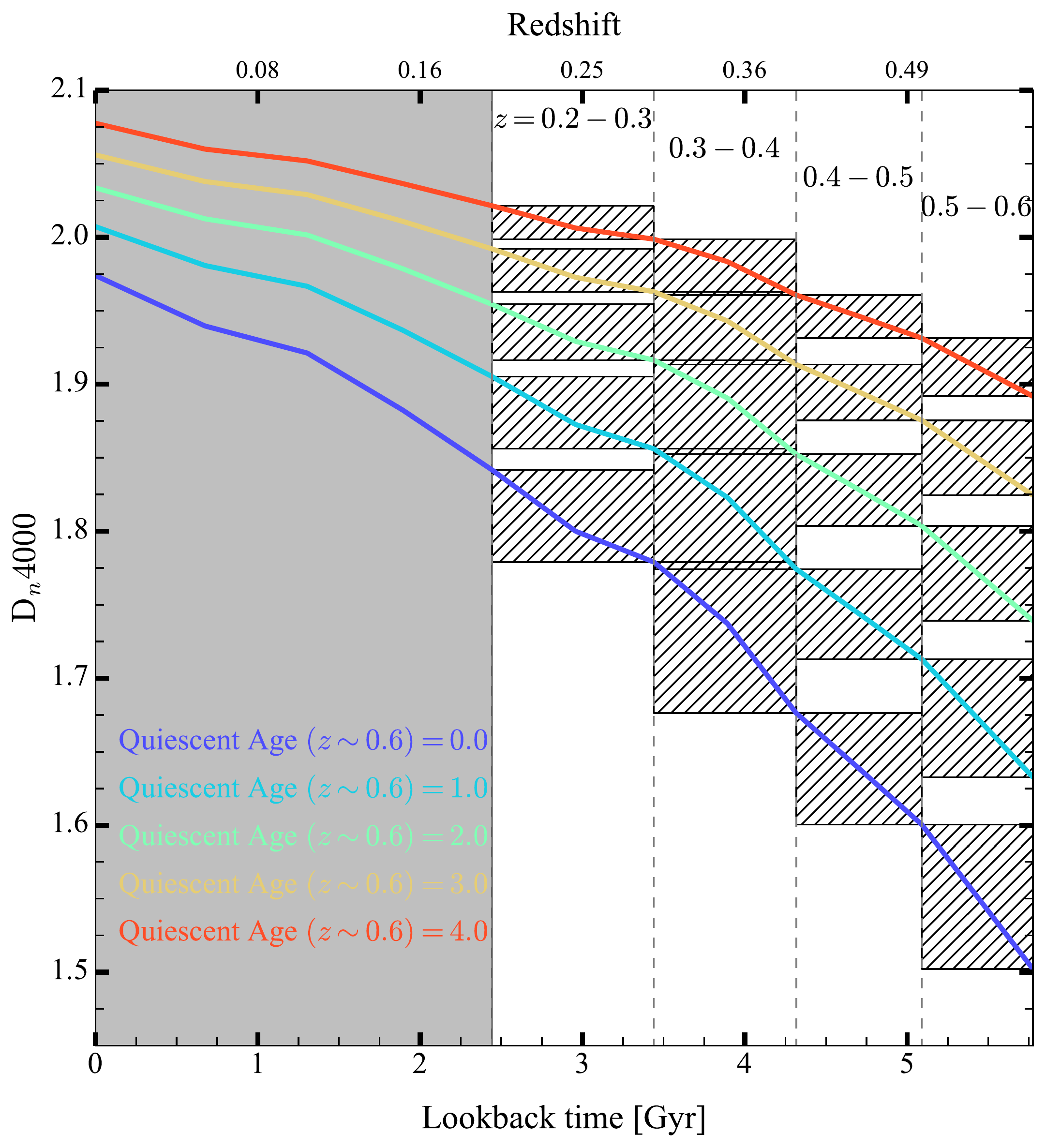}
\caption{$\dn$ as a function of lookback time for a model galaxy that has been quiescent for 0 (purple) to 4 (red) Gyrs at redshift $z \sim 0.6$. We use four redshift (lookback time) bins (dashed gray lines) to explore the evolution of the size -- mass relation of aging quiescent galaxy subpopulations in Figure~\ref{f9}. The hashed bins delineate the subsamples we use to trace the size -- mass relation for the aging quiescent population.
\label{f7}}
\end{centering}
\end{figure}

Figure \ref{f7} shows the evolution of $\dn$ for objects with 5 different values of $\dn$ at $z \sim 0.6$, the limit of the redshift survey. These $\dn$ values correspond to periods of 0 (purple curve) to 4 Gyr (red curve) since cessation of star formation. 

We compute $\dn$ from the synthetic spectrum of a quiescent galaxy derived from the Flexible Stellar Population Synthesis code \citep[FSPS;][]{conroy2009,Conroy2010}. The model galaxy has solar metallicity and constant star formation rate for 1 Gyr. 

The value of $\dn$ has some dependence on metallicity \citep{Poggianti97,Balogh99,Bruzual03}. \citet{Zahid19} investigate the sensitivity to metallicity of evolutionary tracks like the ones in Figure \ref{f7}. They run an FSPS model with constant star formation rate for 1 Gyr and twice solar metallicity. Their results are insensitive to this choice relative to the assumption of solar metallicity. Their Appendix provides details and demonstrates that statistical errors rather than the potential impact of metallicity are the main contributor to uncertainty in the results of this approach to analyzing the data.

\begin{figure*}
\begin{centering}
\includegraphics[width=0.9\textwidth]{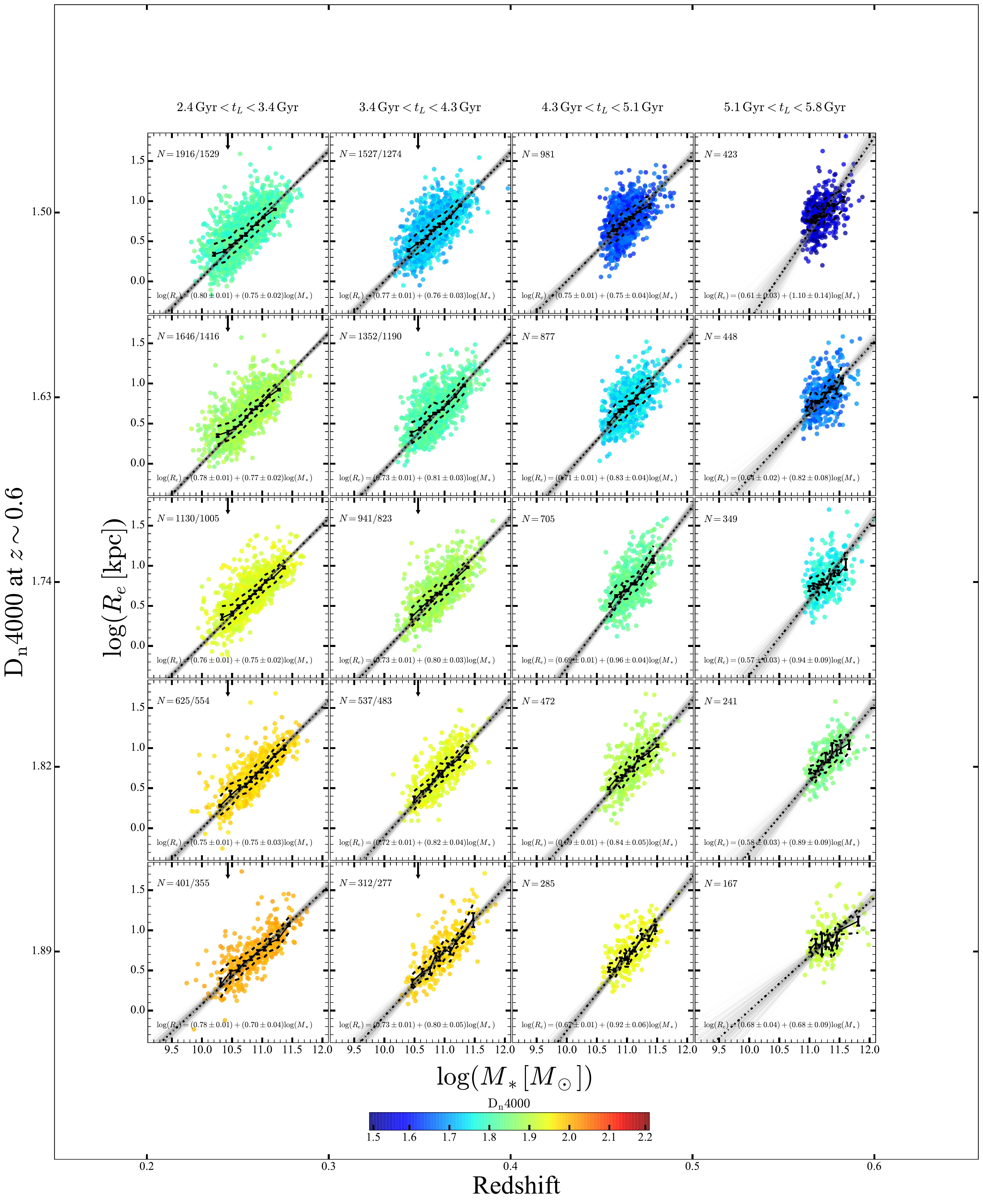}
\caption{Size-stellar mass relation for the descendants of quiescent galaxies with $\dn$ ($z \sim 0.6$) (left-hand scale) implying a quiescent age at $z = 0.6$ ranging from 0 to 4~Gyrs in 1~Gyr steps (top to bottom). The four columns show resident subpopulations aging according to the model in Figure~\ref{f7} as the redshift (lower scale) changes. In each panel the solid black line connects the median sizes in 10 equally populated mass bins. The errors are bootstrapped. The region between the broken dashed lines shows the intrinsic scatter and includes 68\% of the sample in each mass bin. The central black dash-dotted line and gray region trace the best-fit linear relation between stellar mass and size along with the associated uncertainties. The equation for the best-fit size-stellar mass relation in the legend of each cell includes the effective (half-light) radius $R_e$ measured in kpc and stellar mass $M_\ast$ in units of $10^{11}~M_\sun$. Vertical black arrows indicate the pivot stellar mass from Figure~\ref{f5}.
\label{f8}}
\end{centering}
\end{figure*}

In five rows of four panels each, Figure \ref{f8} delineates the evolution of quiescent objects with ages $\tau = [0, 1, 2, 3, 4]$ Gyr at $z \sim 0.6$ (corresponding to the curves and bins in Figure \ref{f7}) by using $\dn$ to identify the descendants of these objects at successive redshift (or, equivalently, look back time) intervals covering the range $0.2 < z < 0.5$. We use the model in Figure \ref{f7} to identify the descendants based on their measured $\dn$. This procedure removes all of the quiescent objects that only enter the quiescent population at redshift $z < 0.5$. The hashed boxes in Figure \ref{f7} delineate the samples of quiescents that we use to trace the evolution of the population in each redshift bin. The $\dn$ limits of the bins are defined by the values where of the model evolutionary curves at the redshift limits of successive bins. In other words, the model defines the binning in $\dn$. We can then explore the dependence of the size -- mass relation on redshift for quiescent objects in these bins.

In each row of Figure \ref{f8}, the slopes (each panel shows the size -- mass relation) of the relations remain roughly constant. The slope is poorly constrained at the highest redshift where the sample is the smallest. The steady increase in the normalization of the relation with decreasing redshift (left to right) along each row demonstrates the size growth of the resident population.

\begin{figure*}
\begin{centering}
\includegraphics[width=0.9\textwidth]{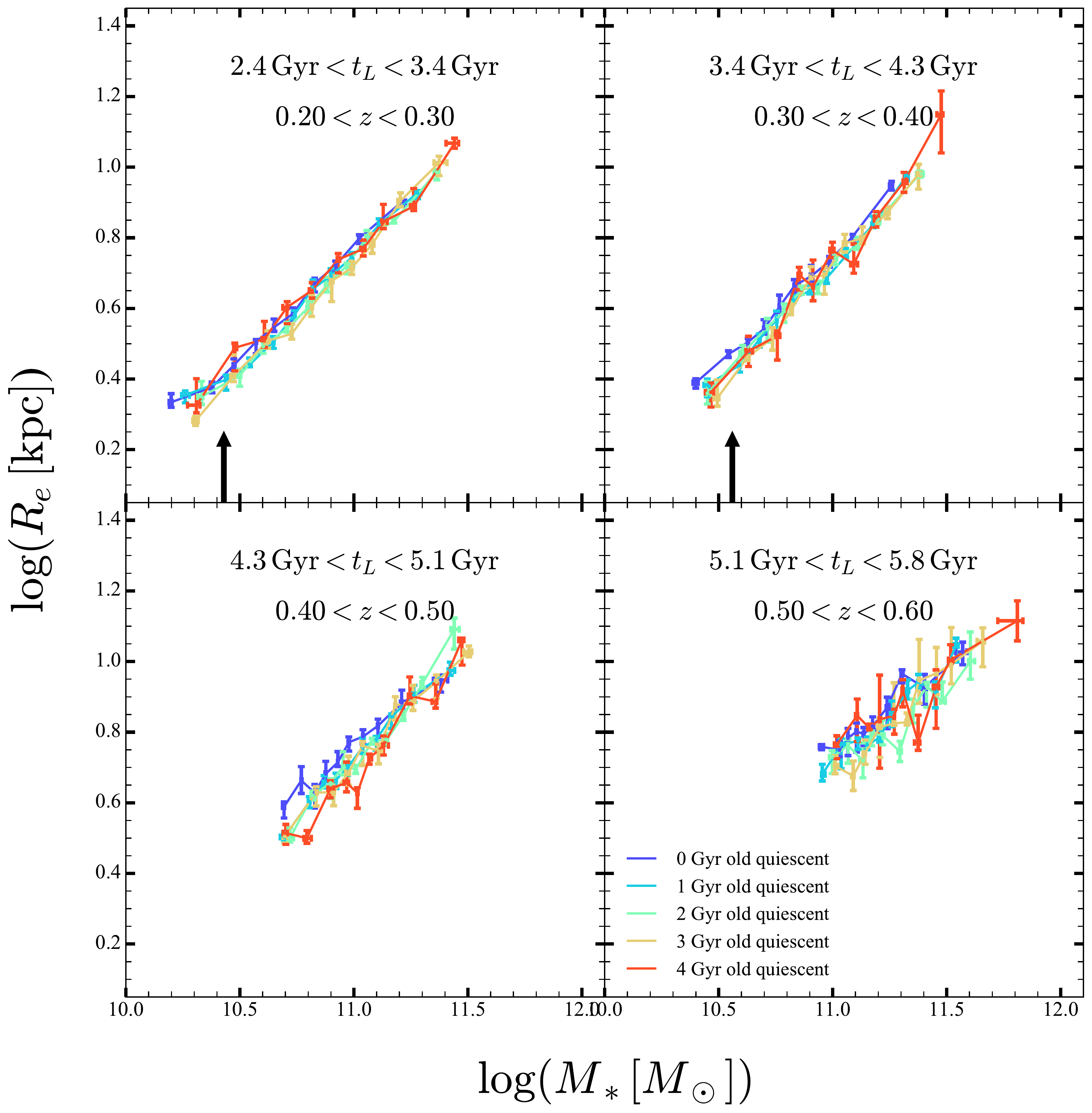}
\caption{Median half-light radius along the major axis as a function of median stellar mass in 10 equally populated mass bins for quiescent HectoMAP galaxies in four redshift/lookback time intervals. The colors are as in Figure \ref{f7} and denote galaxies that have been quiescent for times ranging from 0 to 4 Gyrs at $z \sim 0.6$. We track quiescent aging over the $0.2 < z < 0.6$ redshift interval (Figures \ref{f7} and ~\ref{f8}). Errors in the median values are bootstrapped. The trend in galaxy size with stellar mass is essentially identical for all quiescent resident populations regardless of the $\dn$ (or age since quiescence) at $z \sim 0.6$. Vertical arrows indicate pivot stellar masses from Figure~\ref{f5}.
\label{f9}}
\end{centering}
\end{figure*}

To highlight these results, Figure \ref{f9} shows a superposition of panels in intervals of fixed lookback time from Figure \ref{f8}. The result is striking: for the entire range of ages for the parent $z \sim 0.6$ population, the relations for their descendants are essentially coincident at every lookback time probed by HectoMAP. The size growth of the resident population is independent of the galaxy age at $0.5 < z < 0.6$ once the impact of newcomers to quiescence is removed at lower redshifts ($z < 0.5$). In other words, the growth of the resident population depends only on the lookback time (or, equivalently, change in redshift) since the objects joined the quiescent population.

\begin{figure*}
\begin{centering}
\includegraphics[width=0.9\textwidth]{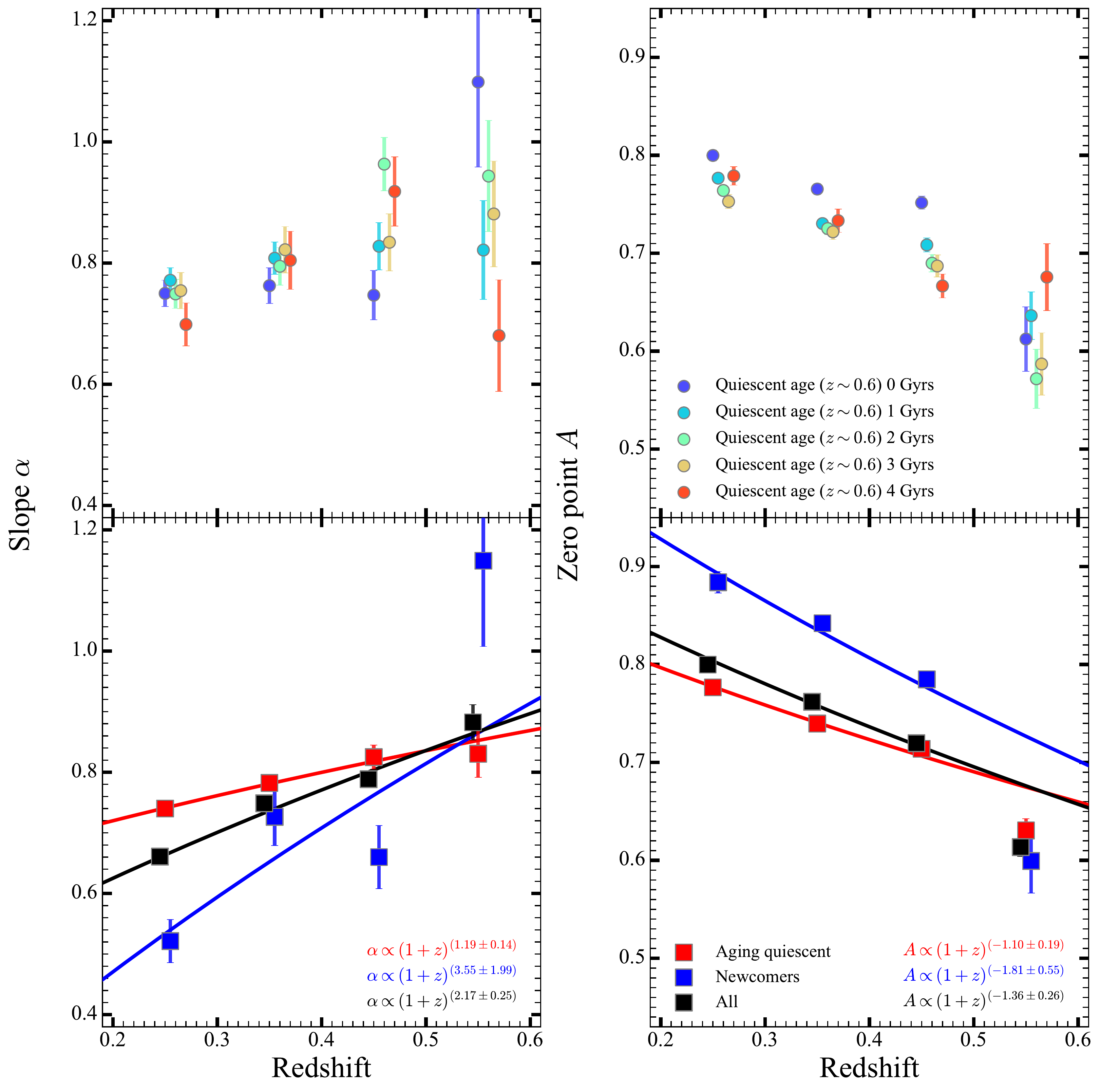}
\caption{{\it Top panels:} Evolution in the slope $\alpha$ (left panel) and zero point $A$ (right panel) for the size-stellar mass relations $\log(R_e/1\, \mathrm{kpc})=A+\alpha\times\log(M_\ast/10^{11}\, M_\sun)$ (same colors as Figures~\ref{f8} and~\ref{f9}. The change in zero point demonstrates the significant size growth. {\it Bottom panels:} Evolution in the parameters of the size-stellar mass relation for newcomers (blue), aging quiescents from Figure \ref{f8} (red) and the entire quiescent sample from Figure \ref{f5} (black). The colored lines show the error-weighted best fits for the evolution for the slope and zero point of the size -- mass relation as a function of redshift.
\label{f10}}
\end{centering}
\end{figure*}

The upper panels of Figure \ref{f10} enhance the comparison in Figure \ref{f9} by displaying the dependence of the parameters of the size -- mass relation on redshift for the $\dn$ subsamples in Figure \ref{f8}. The slope (left panel) is insensitive to redshift
within the error but the size (right panel) grows significantly as the universe ages. As suggested in Figure~\ref{f8}, the behavior of the size -- mass relation is remarkably similar for all of the descendant subsamples. In other words, for galaxies that were quiescent at $z \sim 0.6$, size growth depends only on the redshift and is independent of the time when the object became quiescent.

The lower panels of Figure \ref{f10} follow the evolution of the size -- mass relation parameters for several subsamples of the total sample in Figure \ref{f5}: the newcomers (blue), the aging quiescents of Figure \ref{f8} (red), and the entire quiescent sample (Figure \ref{f5}; black). Table~\ref{tab2} lists the parameters of the size -- mass relations in these two panels. The entire sample of HectoMAP quiescent objects (black squares) includes newcomers, the population of residents aging from $z = 0.6$ (Figure \ref{f8}), and the descendants of newcomers that join the quiescent population at redshifts $\lesssim 0.5$. The newcomers are $\sim15\%$ of the total population and the aging quiescent population from Figure~\ref{f8} constitute $\sim 40\%$ of the quiescent sample. 

For all of the samples highlighted in the lower panels of Figure \ref{f10}, the slope of the size -- mass relation decreases as the universe ages (lower left) and the zero point of the relation increases (lower right). The fits listed in the legend take the errors in the individual points into account. 

The trend in the zero point $A$ of the size -- mass relation with redshift represents the characteristic size growth of a $M_\ast=10^{11}\, M_\sun$ galaxy. For the entire HectoMAP quiescent sample, the zero point trend echoes previous results. The change in slope of the relation ($\alpha$) differs from the constant value (within the uncertainties) reported in some earlier studies (Section~\ref{SM}). More recent studies, however, reveal a increase in $\alpha$ with redshift between $z\sim0.3$ and $z\sim1.1$ \citep{Kawin21,Hamadouche22}. We discuss the trend in $\alpha$ for the HectoMAP quiescent sample further in Section~\ref{limit}. 

Both the slope and zero point of the size -- mass relation show the steepest change for the newcomers. The slope and zero point change less steeply for the aging quiescents of Figure \ref{f8} than for the entire quiescent sample of Figure \ref{f5}. The combination of newcomers and their descendants and related `downsizing in time' \citep{Cowie96,Neistein06} account for this difference (e.g., Figures 8 and 9 of \citet{Haines17} illustrate the change with redshift in the distribution of star forming and quiescent galaxies within the size -- mass plane between $0.5 < z < 1$ and $z \sim 0$). At fixed stellar mass newcomers and their descendants are larger than the typical resident objects that evolve as quiescent galaxies from $z = 0.6$. These younger objects produce a shallower slope and a steeper increase in the zero point. Quantitatively, the characteristic size of an $M_\ast=10^{11}\, M_\sun$ quiescent galaxy in the full mass limited sample increases 48\% between $z = 0.6$ and $z = 0.2$. For the aging resident population that is already quiescent at $z = 0.6$ the fractional increase in size is 37\% over the same redshift range. The contribution of newcomers and their descendants causes  characteristic size growth of
48\% rather than by 37\% from $z = 0.2$ to $z = 0.6$.

\begin{deluxetable*}{l  c  c c}\label{tab2}
\tabletypesize{\small}
\tablecaption{Parameters of the size - mass relation $\log(R_e/1\, \mathrm{kpc})=A+\alpha\times\log(M_\ast/10^{11}\, M_\sun)$ for the subsamples of quiescent HectoMAP population}
\tablewidth{0pt}
\tablehead{
\colhead{Sample} & {Redshift range}  & {Slope $\alpha$} &  {Zero point $A$}}
\startdata 
&{$0.2<z<0.3$}&& \\
\hline
{Quiescent sample}\tablenotemark{a} & &$0.661\pm0.009$  &$0.799\pm0.003$  \\
{Newcomers}\tablenotemark{b} &&$0.500\pm0.035$   &$0.887\pm0.011$ \\
{Aging population}\tablenotemark{c}&&$0.740\pm0.011$ & $0.777\pm0.003$ \\
\hline
&{ $0.3<z<0.4$} &&\\
\hline
{Quiescent sample}\tablenotemark{a} &&$0.749\pm0.011$ &$0.762\pm0.020$  \\
{Newcomers}\tablenotemark{b} && $0.722\pm0.047$ & $0.842\pm0.008$ \\
{Aging population}\tablenotemark{c} && $0.782\pm0.014$ &  $0.740\pm0.003$ \\
\hline
&{ $0.4<z<0.5$}&&\\
\hline
{Quiescent sample}\tablenotemark{a} &&$0.788\pm0.014$  & $0.720\pm0.003$ \\
{Newcomers}\tablenotemark{b} && $0.656\pm0.052$ & $0.788\pm0.007$ \\
{Aging population}\tablenotemark{c} && $0.825\pm0.020$ & $0.714\pm0.004$  \\
\hline
&{ $0.5<z<0.6$}&& \\
\hline
{Quiescent sample}\tablenotemark{a} && $0.882\pm0.029$   & $0.614\pm0.010$ \\
{Newcomers}\tablenotemark{b} && $1.089\pm0.133$ & $0.616\pm0.030$ \\
{Aging population}\tablenotemark{c} && $0.830\pm0.039$  & $0.631\pm0.012$  \\
\enddata
\tablenotetext{a}{Galaxies with D$_n4000>1.5$ from Figure~\ref{f5}.}
\tablenotetext{b}{Galaxies entering quiescent population ($1.5<\mathrm{D}_n4000<1.6$).}
\tablenotetext{c}{Subpopulation of galaxies that have been quiescent for $0-4$~Gyr at $z\sim0.6$ and their descendants (Figures~\ref{f7} and~\ref{f8}).}
\end{deluxetable*}

\begin{figure*}
\begin{centering}
\includegraphics[width=0.9\textwidth]{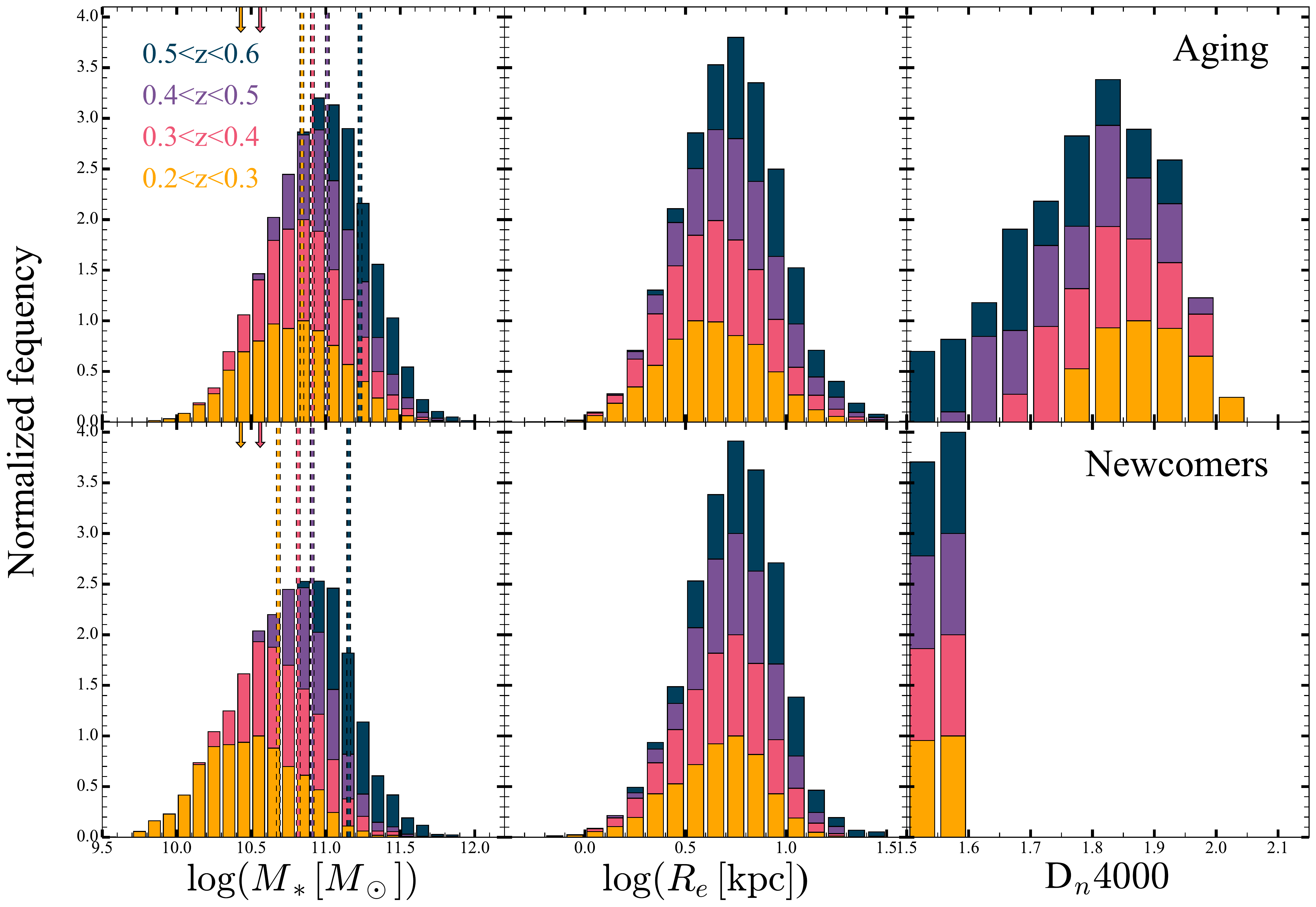}
\caption{Distributions of stellar mass (left panels), size (central panels), and $\dn$ index (right panels) for two subpopulations of intermediate-redshift quiescent galaxies: the aging population (with $\dn$ values defined in Figure~\ref{f7}, top panels) and newcomers (with $1.5 < \dn <1.6$, bottom panels). Different colors correspond to the four redshift bins within the $0.2 < z < 0.6$ range of the HectoMAP survey. The histograms are stacked. Each histogram is normalized to the most populated bin. Vertical arrows show the pivot stellar masses from Figure~\ref{f5}. Dashed lines show the median stellar mass (above the pivot point at lower redshift) in each redshift bin. Comparison of median masses in each redshift bin of the aging and newcomer populations reveals the more prominent downsizing of the newcomers.} 
\label{f11}
\end{centering}
\end{figure*}

The histograms in Figure \ref{f11} underscore the results in the lower panels of Figure \ref{f10}. They show the distributions of stellar mass, size, and $\dn$ for the aging population (top row) and for the newcomers (bottom row). The colored bars in the stacked histograms denote the discrete redshift slices. The shifts of the newcomers toward lower stellar mass (colored dashed vertical lines indicate the median mass for each redshift bin) and larger sizes are evident in each redshift bin. The narrow $\dn$ distribution in the lower right panel defines the sample of newcomers at each redshift.

Downsizing is a larger effect in the newcomer population (lower left) relative to the aging residents (upper left). This effect is expected, but its amplitude has not been measured previously in this redshift range. From $z \sim 0.55$ to $z \sim 0.25$ the change in the median stellar mass for aging population is 0.1~dex smaller than for the population of newcomers. 

\begin{figure}
\begin{centering}
\includegraphics[width=0.45\textwidth]{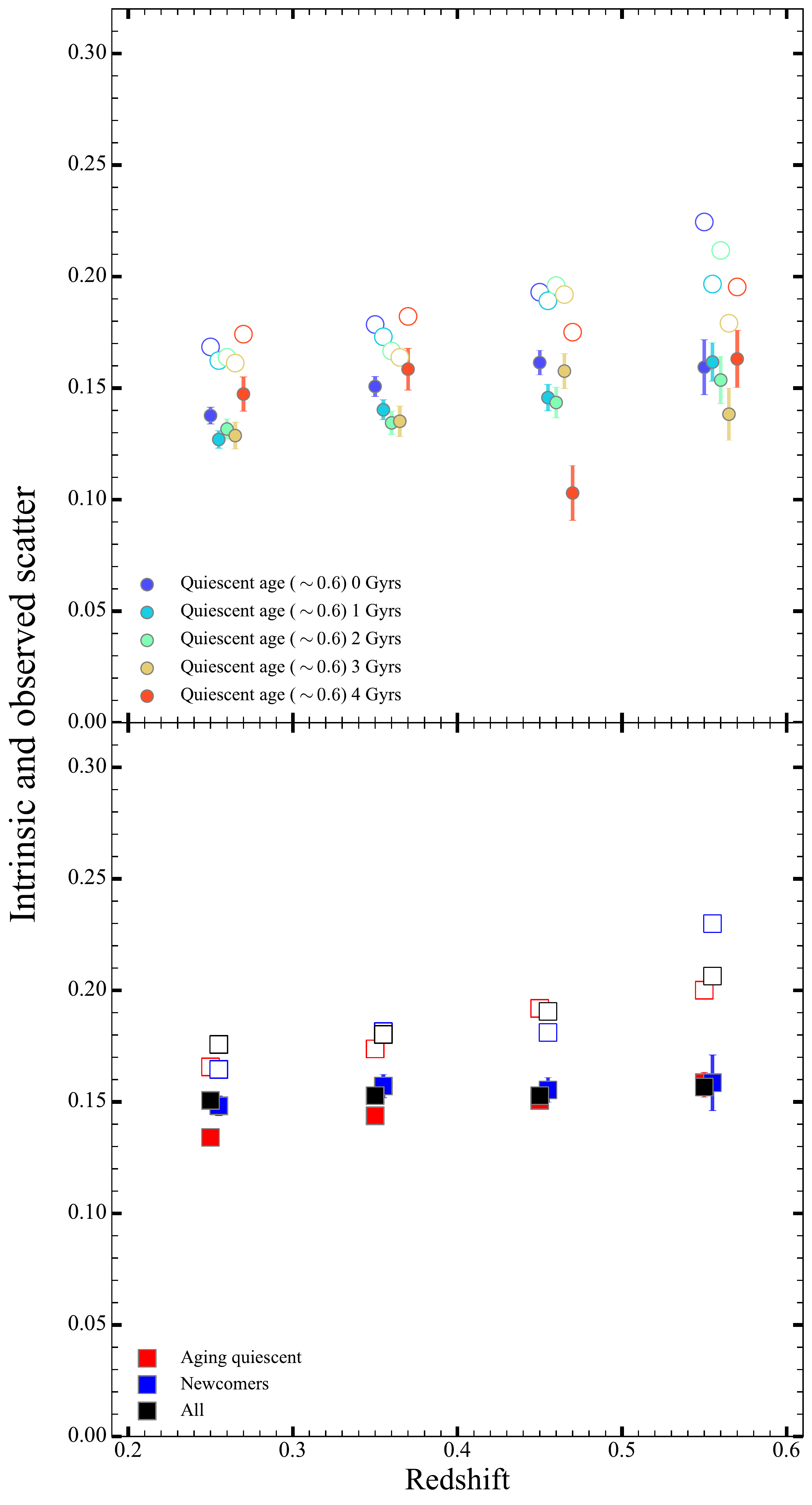}
\caption{{\it Top panel:} Intrinsic scatter (i.e., $\sigma$ of the fitted size -- mass relation $\log(R_e)=A+\alpha\times\log(M_\ast)+\sigma$; filled circles) and the standard deviation in measured sizes after removal of the best-fit  relation (observed scatter, open circles) for the aging sub-populations from Figure~\ref{f8}. {\it Bottom panel}: The intrinsic and observed scatter for the three quiescent samples from the bottom panels of Figure~\ref{f10}. 
\label{f12}}
\end{centering}
\end{figure}

A number of studies investigate the scatter around the size -- mass relation as a measure of the characteristic evolution of the quiescent population \citep[e.g.,][]{Shen03,vanderWel14,Mowla19,Yang21}. Figure \ref{f12} examines the behavior of both the intrinsic and observed scatter for the size -- mass relations of the populations aging from $z = 0.6$ (upper panels) and the total aging sample, newcomers, and the total quiescent sample (lower panels). The intrinsic scatter (the $\sigma$ of the fitted size -- mass relations) is insensitive to redshift, in agreement with \citet{Shen03} and \citet{vanderWel14}. The observed scatter (the standard deviation in observed sizes after removal of the fitted size -- mass relation) always exceeds the intrinsic scatter as expected. The observed scatter increases mildly with redshift due to the subtle increase $\dn$ measurement uncertainty for $z > 0.3$ \citep{Damjanov22}.

The large HectoMAP quiescent sample enables the derivation of robust parameters that characterize the size -- mass relation for a range of subsamples of the data. These subsamples encompass the newcomers and the full mass-limited quiescent sample. We next explore the size -- mass relations for the HectoMAP quiescent subsamples relative to the minor merger model \citep{Naab09,Bezanson09,Hopkins10a} that could drive the growth of quiescent objects in the redshift range $0.2 < z < 0.6$.

\section{Discussion}
\label{discussion}
\begin{figure*}
\begin{centering}
\includegraphics[width=0.9\textwidth]{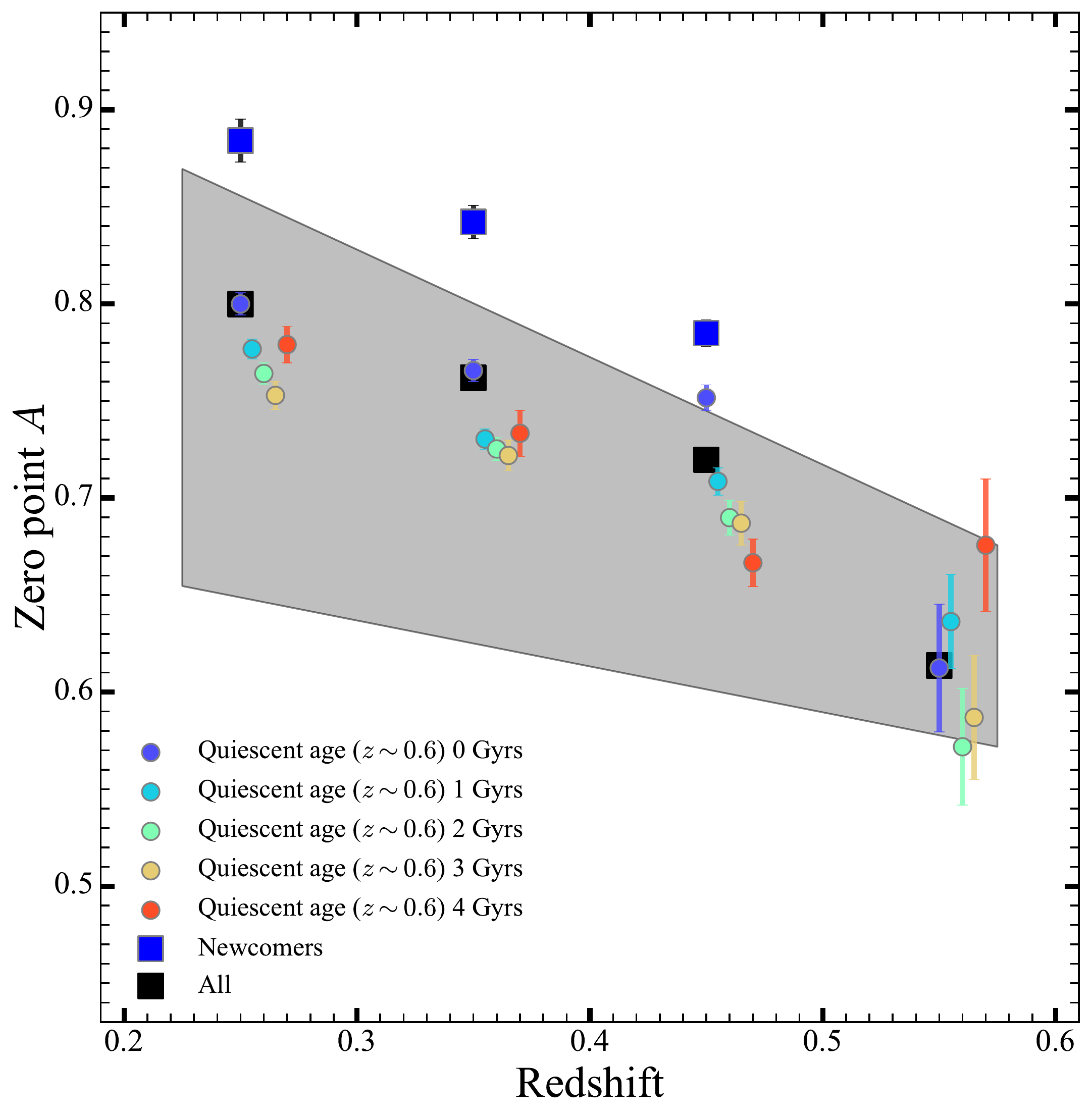}
\caption{Evolution in the characteristic size (the zero point, A) of an $M_\ast=10^{11}\, M_\sun$ for aging quiescent galaxy samples (colored dots; upper right panel of Figure~\ref{f10}). The gray shaded area highlights the range in size growth predicted by \citet{Naab09} minor merger models. The blue squares show the more dramatic change in the characteristic size of a newcomer  $M_\ast=10^{11}\, M_\sun$ galaxy with $1.5<\mathrm{D}_n4000<1.6$ (blue squares in the lower right panel of Figure \ref{f10}). The black squares show the behavior of the characteristic $R_e$ for the entire mass-complete quiescent sample as do black squares in Figure \ref{f10}. The aging quiescent population (not the less abundant newcomers) dominates the overall quiescent size growth.}
\label{f13}
\end{centering}
\end{figure*}

\subsection {Consistency with Minor Merger Driven Growth at $0.2 < z < 0.6$}\label{merge}

HectoMAP provides a dense sampling of the size evolution of the quiescent population for $0.2< z < 0.6$. Spectroscopy provides the redshift and the age indicator $\dn$; HSC images provide the sizes. The density of the sample and the inclusion of a population age indicator enable study of size growth for populations that have different ages at the initial redshift $z \sim 0.6$. The data also enable removal of newcomers from the aging quiescent sample. HectoMAP shows that the growth of the aging quiescent population is independent of the population age.

In the emerging consensus view, major mergers probably play a significant role in the size evolution of the quiescent population at $z \gtrsim 1$, but at lower redshift, minor mergers dominate \citep[e.g.,][]{Trujillo11,Newman12, Lopez-Sanjuan12,McLure13,Conselice14,Greene15,Oh17,Hill17,Matharu19,Zahid19,Mosleh20}. In the redshift range $0.2 < z < 0.6$ covered by HectoMAP several lines of evidence already support the dominance of minor mergers. \citet[][Table 4]{Lotz11} provide a rough estimate of the relative numbers of major and minor mergers per $M_\ast>10^{10}\, M_\sun$ galaxy at $z < 1.5$. Over the redshift range of our survey each galaxy experiences about 1 minor merger; major mergers are a factor of 4-5 less common. The HectoMAP data enable further examination of the consistency between the minor merger driven growth paradigm and the data.

To compare the data with a model for minor merger driven size evolution, the gray shades area in Figure \ref{f13} highlights the range of  theoretical predictions based on the models of \citet{Naab09}. Section 2 of their paper describes predictions based on the virial theorem and energy conservation. To compare the model with the data we need an estimate of the minor merger rate per Gyr. \citet{Lotz11} summarize a range of observational studies in their Table 4. For the HectoMAP redshift interval, the \citet{Lotz11} typical minor merger rate estimate per galaxy is $\sim0.3$~Gyr$^{-1}$ or, equivalently, $\sim 1$ merger in the redshift range. For these minor mergers a reasonable mass ratio for the merging systems is $0.1-0.25$.

The gray shaded region in Figure \ref{f13} covers the range of \citet{Naab09} model predictions (their equation 4). The underlying model assumes that the initial and final systems are in virial equilibrium and that energy is conserved during the minor merger.

The lower limit of the gray region in Figure \ref{f13} is for an accreted/central mass ratio \citep[Eq. 4 of][]{Naab09}, $\eta =0.1$, and the upper limit is for $\eta=0.25$. We assume that $\epsilon$, the ratio between the internal velocity dispersion of the accreted material and the internal velocity dispersion of the original, central galaxy is $\ll 1$; we thus neglect $\epsilon$.

As an interesting counterpoint, Figure \ref{f13} shows the remarkably steep dependence of the size -- mass relation zero point on redshift for newcomers (blue squares). The apparent size evolution of this population is obviously steeper than the minor merger model prediction. Of course, the properties of these populations should not be minor merger driven. The change in size results from the lower stellar mass, larger objects that are just entering quiescence as the universe ages (Figure \ref{f11}). 

For the resident population aging from $z \simeq 0.6$ (colored circles), the  complete overlap between the \citet{Naab09} model predictions and the data 
supports the conjecture that minor mergers are the main driver of the size growth of the aging quiescent population. The black squares show the evolution of the total quiescent population which includes this aging population along with objects that first become quiescent at $z \lesssim 0.5$. As demonstrated by the analysis accompanying Figure \ref{f10} (Section~\ref{reside}), inclusion of newcomers in the population does not change the size evolution enough to depart from the minor merger model predictions. As we note in Section~\ref{reside}, only 15\% of the total population are newcomers and they change the size growth from $z\sim0.6$ to $z\sim0.2$ by $\sim11\%$. Recently, \citet{Hamadouche22} report a similar percentage of newcomers and demonstrate  the dominant effect of minor mergers based on the observed size, stellar mass, and D$_n4000$ of quiescent galaxies in the redshift range $0.7\lesssim z\lesssim1.1$.

The lack of dependence of size evolution of resident descendants on the quiescent age at $z = 0.6$ (colored circles) may also be a signature of minor merger driven growth. These mergers deposit material primarily in the outer regions of the galaxy \citep{Bezanson09,Naab09,Hopkins09a,Oser10} and they are thus unlikely to affect the central stellar population age or, equivalently, the $\dn$ measured in the Hectospec 1.5$^{\prime\prime}$ fiber aperture. The central population age thus can be consistent with passive evolution models even though there is size growth \citep{Whitney21}. Spatially resolved spectroscopy in larger apertures might test this conjecture by revealing the younger stellar populations associated with merger events. 

\subsection{Limitations}\label{limit}

The large complete HectoMAP sample of quiescent galaxies traces the structural evolution of these objects at $z<0.6$ segregated by the average age of their stellar population. The uncertainties in the $\dn$ index, a proxy for the average stellar population age, have no significant effect on the analysis (Appendix~\ref{appendix1}). 

As in any magnitude-limited survey, the $r=21.3$ limit translates to a stellar mass limit that is a function of redshift (Section~\ref{pop}, Figure~\ref{f1}). In the highest redshift bin ($0.5 < z < 0.6$), the mass-complete sample covers stellar masses $M_\ast > 10^{11}\, M_\sun$. In the lowest redshift bin ($0.2 < z < 0.3$) the limiting mass of the complete sample reaches a limiting stellar mass of $\lesssim10^{10}\, M_\sun$. 

By probing lower-mass quiescent systems (and thus a larger dynamic range of stellar masses) with decreasing redshift we demonstrate the dependence of the size -- mass relation on $\dn$ (Figure~\ref{f5}). Including of lower stellar mass, generally younger galaxies, changes  the stellar mass range accessed as a function of redshift and makes the slope of the size -- mass relation appear shallower at lower redshift (Figure~\ref{f10}). 

The change in the slope of the size -- mass relation is most dramatic for the newcomers (from $\alpha (z\sim0.55) \sim 1$ to $\alpha (z\sim0.25) \sim 0.5$, Table~\ref{tab2}) because lower mass galaxies at lower redshift are also younger. The best-fit model for the relation between $\alpha$ and redshift for newcomers from the lower left panel of Figure~\ref{f10} suggests a factor of $\gtrsim2.7$ decrease over the redshift interval of our survey (or $\sim3.3$~Gyr of cosmic time). The slope of the relation decreases at a significantly slower rate for quiescent galaxies aging from $z \sim 0.6$ (from $\alpha(z\sim0.55) \sim 0.85$ to $\alpha(z\sim0.25) \sim 0.75$). The best-fit evolutionary model for this population (Figure~\ref{f10}) suggests a decrease in $\alpha$ of only 30\% over the same redshift interval. The descendants of galaxies with $\dn > 1.5$ at $z \sim 0.6$ are among the most massive galaxies in each redshift bin and thus they are less susceptible to the effects of the broader stellar mass range sampled  with decreasing redshift.

The evolution of the zero point (characteristic size) for an $M_\ast=10^{11}\, M_\sun$ quiescent galaxy, is robust to changes in the range stellar mass range sampled as a function of redshift; the highest-redshift points in the lower right panel of Figure~\ref{f10} only increase the uncertainty in the increase of characteristic size with redshift. Deeper spectroscopic surveys that explore a broad and fixed stellar mass range as a function of redshift will reduce these systematic issues and will enable a more robust probe of the evolution in the slope of the quiescent size -- mass relation.

The evolution in the characteristic size of quiescent objects is based on the change in the {\it observed light profiles} of the objects. In studies that use spatially resolved spectral energy distribution (SED) fitting and/or empirical relationship between the mass-to-light ratio $M/L$ and observed color \citep[e.g.,][]{Szomoru13,Chan16,Suess19,Mosleh20,Miller22} the inferred mass-weighted size (or half-mass size $R_{mass}$) of galaxies is consistently smaller than the size based on the observed light profiles (i.e., the half-light radius $R_e$ we measure). These studies rely on space-based observations ({\it Hubble Space Telescope} or, very recently, {\it James Webb Space Telescope}, \citealt{Suess22}) and almost all of them target high-redshift ($z>1$). The results range from constant $R_{mass}/R_e$ ratios, that do not alter the evolutionary trend in characteristic size, to a reduced growth of half-mass radius with respect to the half-light size due to the trend in $R_{mass}/R_e$ with redshift \citep[for more details, see][]{Miller22}. On the other hand, a recent comparison among different galaxy size estimates based on mock images of $z=0.1$ galaxies in the EAGLE simulations demonstrates that the light-weighted structural properties of quiescent galaxies we use here are a good proxy for their mass-weighted properties \citep{deGraff22}.

With its large-area coverage and dense spectroscopy of quiescent systems at $z < 0.6$, HectoMAP has the potential to resolve the tension between results based on various half-mass size estimates. However, the ground-based HSC SSP imaging of the region is not adequate to this task. The resolving power of the HSC SSP images in $grizy$ (quantified by the size of the PSF) varies significantly across the field and among filters\footnote{\url{https://hsc-release.mtk.nao.ac.jp/doc/index.php/quality-assurance-plots__pdr3/\#wide_hectomap1_QA_calPsfDiff.png-Shape}}$^{,}$\footnote{\url{https://hsc-release.mtk.nao.ac.jp/doc/index.php/quality-assurance-plots__pdr3/\#wide_hectomap2_QA_calPsfDiff.png-Shape}}. Robust estimates of HectoMAP galaxy sizes from spatially resolved stellar mass maps require multi-wavelength imaging with uniformly high (diffraction-limited) resolution over the full 55~deg$^2$ area of the survey. A photometric dataset of this quality may only become available with future space missions  \citep[e.g., Euclid;][]{Amiaux12}. 

Future deep and dense spectroscopic surveys with the next-generation instruments (DESI, \citealt{Zhou22}; MOONS, \citealt{Taylor18}; PFS, \citealt{Greene22}) will reach beyond $z\sim1$ enabling an extensive exploration of the relative roles of major and minor mergers in the evolution of the quiescent population. We show that, in agreement with the broad consensus view, minor mergers are the dominant drivers of the size growth in quiescent population at $z < 0.6$ \citep[e.g.,][]{Ownsworth14}. In this redshift interval, minor mergers are several times more frequent than major mergers \citep{Lotz11}. Future spectroscopic surveys will probe the higher redshift, earlier epoch where major mergers may become more dominant drivers of galaxy growth through mass accretion \citep[][and references therein]{Mantha18}. Major mergers have a stronger effect on the  central morphology, dynamics, and stellar population of quiescent galaxies \citep[e.g.,][]{Toomre77,White78,Hopkins09b,Bois11} that should be directly observable at $ z \gtrsim 1$.

\section{Conclusion}\label{con}

Large, complete, mass limited samples of quiescent galaxies provide an increasingly detailed picture of their evolution. The HectoMAP survey combines MMT spectroscopy and Subaru HSC imaging to yield a mass limited sample of 30,231 quiescent objects spanning the redshift range $0.2 < z < 0.6$. The spectroscopy provides a redshift along with the age indicator, $\dn$. HSC SSP $i-$band imaging currently provides half-light radii for quiescent objects in 80\% of the total area covered by HectoMAP. We derive stellar masses from SDSS photometry. The HectoMAP quiescent sample is 6.6 times the size of the largest comparable sample available previously. We explore the size -- mass relation for the entire dataset and for well-defined quiescent subpopulations.

The global size -- mass relations for the HectoMAP survey underscore the known increase in half-light radius with decreasing redshift. The large size of the data set along with the availability of the age indicator, $\dn$, reveals a dependence of the size -- mass relation on $\dn$ that is increasingly striking as the universe ages. As at $z \sim 0$ \citep[based on SDSS,][]{Zahid17} and $z \sim 0.7$ \citep[based on LEGA-C ,][]{Wu18}, $0.2<z<0.6$ quiescent galaxies with smaller $\dn$ are larger at fixed stellar mass. 

The HectoMAP quiescent sample is large enough to extract several subsamples based on $\dn$ that elucidate aspects of the size evolution of the total quiescent population. We define a sample of newcomers with $1.5 < \dn < 1.6$ at each redshift. These objects have just joined the quiescent population. They constitute $\sim 15$\% of the total HectoMAP quiescent sample. We examine the evolution of this population alone by comparing samples in two redshift bins centered at $z \sim 0.25$ and $z = 0.45$. Over this $\sim 1.5$ Gyr interval, the size growth of the newcomers is $\Delta\log{R_e} = 0.11\pm 0.025$. In the higher redshift bin there is a tail toward small $R_{e}$ that is absent at lower redshift. This comparison is a unique and direct measure of the impact of down-sizing \citep[e.g., Figure~9 of][]{Haines17} in the population entering quiescence.

At the limiting redshift of HectoMAP ($z\sim0.6$), we segregate the quiescent population in 5 bins based on the correspondence between age since the onset of quiescence and $\dn$. We use the FSPS model \citep{Conroy2010} to connect the descendants of these objects in three lower redshift intervals with their progenitors. The objects in the five $z \sim 0.6$ age bins and their descendants constitute 40\% of the total HectoMAP quiescent sample and we identify them as the aging quiescent subsample. Comparison of the size -- mass relations for the five $z = 0.6$ age intervals yields the novel and intriguing result that the evolution of the relation is independent of the quiescent ``age" at $z \sim 0.6$. In other words, the size growth depends only on the time interval where we track the evolution.

We compare the parameters of the size -- mass relations for the newcomers and aging quiescents with the total HectoMAP quiescent population. The size growth is significantly steeper for the newcomers. In the full quiescent sample, the size of a fiducial $10^{11}\, M_\odot$ galaxy increases by 48\% and the size of the aging quiescents increase by 37\%. Thus the newcomers contribute 11\% to the growth. This difference is often called progenitor bias \citep[e.g.,][]{Saglia10}; the large, complete HectoMAP sample provides a direct measure of its impact.

The slope of the size -- mass relation appears to become shallower as the universe ages. The effect is largest for the newcomers. This result may depend at least in part on the change in the range of stellar masses we sample as a function of redshift. At lower redshift the sample includes objects with lower stellar mass where the size -- mass relation is known to flatten. Further exploration of this issue obviously requires a deeper sample that encompasses a broad  and fixed stellar mass range throughout the redshift range.

We compare the suite of HectoMAP size -- mass relations with the widely supported minor merger model for its evolution. The evolution of the newcomers alone is too steep to lie within the model boundaries. However, the contribution of these newcomers to the growth of the entire population has a negligible effect on the overall agreement between the evolution of the quiescent population and the model. 

The HectoMAP quiescent sample demonstrates the power of a large, complete mass-limited quiescent sample for disentangling the impact of different subpopulations on the evolution of the size -- mass relation. Deeper redshift surveys will further constrain the evolution by enabling exploration of large, fixed stellar mass ranges over a sizeable redshift range. At larger redshift, the larger impact of major mergers on growth can also be explored. Ultimately both half-light and half-mass radii for the quiescent population will be derived robustly from space-based photometry. Taken together these future observations will fill out a complete picture of the evolution of the quiescent population.\\

I.D. acknowledges the support of the Canada Research Chair Program and the Natural Sciences and Engineering Research Council of Canada (NSERC, funding reference number RGPIN-2018-05425). J.S. is supported by the CfA Fellowship. Y.U. is supported by the U.S. Department of Energy under contract number DE-AC02-76-SF00515. M.J.G. acknowledges the Smithsonian Institution for support. I.D.A. gratefully acknowledges support from DOE grant DE-SC0010010 and NSF grant AST-2108287. This research has made use of NASA’s Astrophysics Data System Bibliographic Services. 

The Hyper Suprime-Cam (HSC) collaboration includes the astronomical communities of Japan and Taiwan as well as Princeton University. The HSC instrumentation and software were developed by the National Astronomical Observatory of Japan (NAOJ), the Kavli Institute for the Physics and Mathematics of the Universe (Kavli IPMU), the University of Tokyo, the High Energy Accelerator Research Organization (KEK), the Academia Sinica Institute for Astronomy and Astrophysics in Taiwan (ASIAA), and Princeton University.
Funding was contributed by the FIRST program from the Japanese Cabinet Office, the Ministry of Education, Culture, Sports, Science and Technology (MEXT), the Japan Society for the Promotion of Science (JSPS), Japan Science and Technol- ogy Agency (JST), the Toray Science Foundation, NAOJ, Kavli IPMU, KEK, ASIAA, and Princeton University. This paper makes use of software developed for the Large Synoptic Survey Telescope (LSST). We thank the LSST Project for making their code available as free software at http://dm.lsst. org. This paper is based [in part] on data collected at the Subaru Telescope and retrieved from the HSC data archive system, which is operated by Subaru Telescope and Astronomy Data Center (ADC) at National Astronomical Observatory of Japan. Data analysis was in part carried out with the cooperation of Center for Computational Astrophysics (CfCA), National Astronomical Observatory of Japan.

\vspace{5mm}
\facilities{MMT Hectospec, Subaru Hyper Suprime-Cam}

\software{SciPy \citep{2020SciPy-NMeth}, NumPy \citep{2020NumPy-Array}, astropy \citep{Astropy13,Astropy18, Astropy22}, Scikit-learn \citep{pedregosa2011scikit}, PWLF: piecewise linear fitting \citep{pwlf},
          Source Extractor \citep{Bertin96}, PSFEx  \citep{Bertin11}
          }

\appendix
\restartappendixnumbering
\section{The impact of the error in D$_n4000$}\label{appendix1}

To evaluate the effect of spectroscopic measurement uncertainties on the quiescent evolutionary trends, we create a set of ten simulated parent quiecent galaxy samples. We simulate ten full simulated datasets of the 79175 individual D$_{n}4000$ measurements for HectoMAP galaxies within the $i-$band HSC coverage. We draw ten values from a Gaussian probability distribution function (pdf) for each $\dn$ measurement. Each Gaussian is centered on the measured value and has a scale scale equal to the measurement error. Thus the simulation captures the  increase in the $\dn$ error for fainter, higher redshift objects in the  original HectoMAP dataset. 

We then follow the analysis steps throughout this paper based on these simulated parent quiescent samples. We evaluate the distribution of the resulting size - mass relation parameters for all of the quiescent galaxy (sub)sets (newcomers, aging population, and the entire quiescent population) in the four redshift bins from Figures~\ref{f5},~\ref{f8},~and~\ref{f10}.

\begin{figure*}
\begin{centering}
\includegraphics[width=0.9\textwidth]{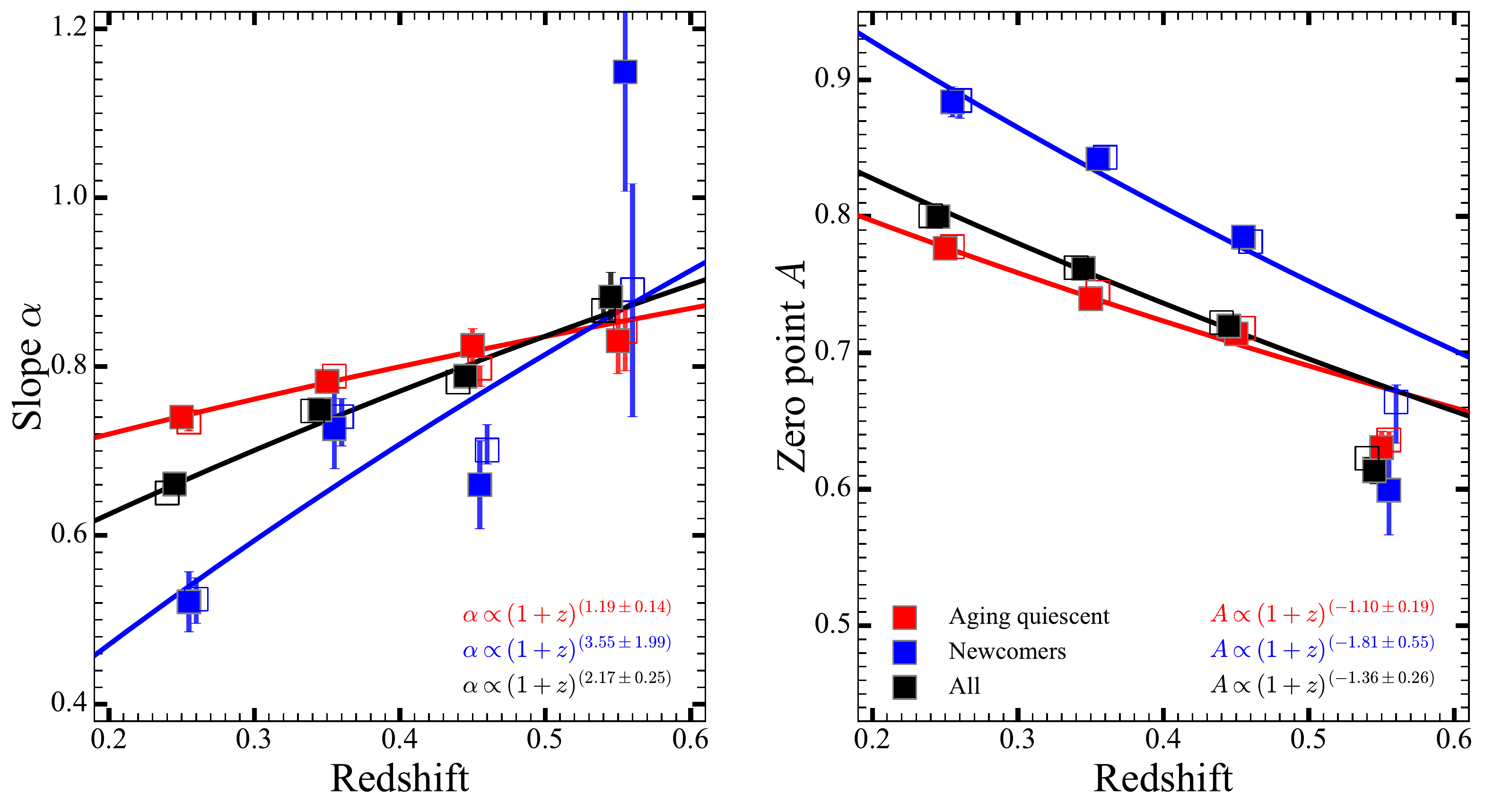}
\caption{Evolution in the size-stellar mass relation for newcomers (blue), aging quiescents (red) and the entire quiescent sample (black) as in Figure~\ref{f10}. The open squares with errorbars show the median values for fitted parameters and their $[16,84]\%$ range based on ten simulated samples with perturbed D$_n4000$ indices. The colored lines show the error-weighted best fits for the evolution for the slope and zero point of the size -- mass relation as a function of redshift from Figure~\ref{f10}. The points based  on the simulated samples follow the evolutionary trends in the slope and zeropoint we derive with the original samples. 
\label{A1}}
\end{centering}
\end{figure*}

Figure~\ref{A1} shows the bottom two panels of Figure~\ref{f10} with additional points (open squares with errorbars) corresponding to the simulation results. Each open square show the median value of the derived size -- mass relation parameter (slope $\alpha$ or zero point $A$). The errorbar shows the $[16, 84]\%$ spread around the median. 

In all redshift bins and in both panels the difference between the $y-$ axis positions of the filled squares (original galaxy samples) and open squares (simulated samples) is minimal and firmly within the uncertainties in parameters from the original galaxy sample. The simulations demonstrate that the observed evolutionary trends are not affected by the spectroscopic measurement uncertainties.

\begin{thebibliography}{}
\expandafter\ifx\csname natexlab\endcsname\relax\def\natexlab#1{#1}\fi
\providecommand{\url}[1]{\href{#1}{#1}}
\providecommand{\dodoi}[1]{doi:~\href{http://doi.org/#1}{\nolinkurl{#1}}}
\providecommand{\doeprint}[1]{\href{http://ascl.net/#1}{\nolinkurl{http://ascl.net/#1}}}
\providecommand{\doarXiv}[1]{\href{https://arxiv.org/abs/#1}{\nolinkurl{https://arxiv.org/abs/#1}}}

\bibitem[{{Ahumada} {et~al.}(2020){Ahumada}, {Prieto}, {Almeida}, {Anders},
  {Anderson}, {Andrews}, {Anguiano}, {Arcodia}, {Armengaud}, {Aubert}, {Avila},
  {Avila-Reese}, {Badenes}, {Balland}, {Barger}, {Barrera-Ballesteros}, {Basu},
  {Bautista}, {Beaton}, {Beers}, {Benavides}, {Bender}, {Bernardi}, {Bershady},
  {Beutler}, {Bidin}, {Bird}, {Bizyaev}, {Blanc}, {Blanton}, {Boquien},
  {Borissova}, {Bovy}, {Brandt}, {Brinkmann}, {Brownstein}, {Bundy}, {Bureau},
  {Burgasser}, {Burtin}, {Cano-D{\'\i}az}, {Capasso}, {Cappellari}, {Carrera},
  {Chabanier}, {Chaplin}, {Chapman}, {Cherinka}, {Chiappini}, {Doohyun Choi},
  {Chojnowski}, {Chung}, {Clerc}, {Coffey}, {Comerford}, {Comparat}, {da
  Costa}, {Cousinou}, {Covey}, {Crane}, {Cunha}, {Ilha}, {Dai}, {Damsted},
  {Darling}, {Davidson}, {Davies}, {Dawson}, {De}, {de la Macorra}, {De Lee},
  {Queiroz}, {Deconto Machado}, {de la Torre}, {Dell'Agli}, {du Mas des
  Bourboux}, {Diamond-Stanic}, {Dillon}, {Donor}, {Drory}, {Duckworth},
  {Dwelly}, {Ebelke}, {Eftekharzadeh}, {Davis Eigenbrot}, {Elsworth},
  {Eracleous}, {Erfanianfar}, {Escoffier}, {Fan}, {Farr},
  {Fern{\'a}ndez-Trincado}, {Feuillet}, {Finoguenov}, {Fofie},
  {Fraser-McKelvie}, {Frinchaboy}, {Fromenteau}, {Fu}, {Galbany}, {Garcia},
  {Garc{\'\i}a-Hern{\'a}ndez}, {Oehmichen}, {Ge}, {Maia}, {Geisler}, {Gelfand},
  {Goddy}, {Gonzalez-Perez}, {Grabowski}, {Green}, {Grier}, {Guo}, {Guy},
  {Harding}, {Hasselquist}, {Hawken}, {Hayes}, {Hearty}, {Hekker}, {Hogg},
  {Holtzman}, {Horta}, {Hou}, {Hsieh}, {Huber}, {Hunt}, {Chitham}, {Imig},
  {Jaber}, {Angel}, {Johnson}, {Jones}, {J{\"o}nsson}, {Jullo}, {Kim},
  {Kinemuchi}, {Kirkpatrick}, {Kite}, {Klaene}, {Kneib}, {Kollmeier}, {Kong},
  {Kounkel}, {Krishnarao}, {Lacerna}, {Lan}, {Lane}, {Law}, {Le Goff}, {Leung},
  {Lewis}, {Li}, {Lian}, {Lin}, {Long}, {Longa-Pe{\~n}a}, {Lundgren}, {Lyke},
  {Ted Mackereth}, {MacLeod}, {Majewski}, {Manchado}, {Maraston}, {Martini},
  {Masseron}, {Masters}, {Mathur}, {McDermid}, {Merloni}, {Merrifield},
  {M{\'e}sz{\'a}ros}, {Miglio}, {Minniti}, {Minsley}, {Miyaji}, {Mohammad},
  {Mosser}, {Mueller}, {Muna}, {Mu{\~n}oz-Guti{\'e}rrez}, {Myers}, {Nadathur},
  {Nair}, {Nandra}, {do Nascimento}, {Nevin}, {Newman}, {Nidever}, {Nitschelm},
  {Noterdaeme}, {O'Connell}, {Olmstead}, {Oravetz}, {Oravetz}, {Osorio},
  {Pace}, {Padilla}, {Palanque-Delabrouille}, {Palicio}, {Pan}, {Pan},
  {Parker}, {Paviot}, {Peirani}, {Ram{\'r}ez}, {Penny}, {Percival},
  {Perez-Fournon}, {P{\'e}rez-R{\`a}fols}, {Petitjean}, {Pieri},
  {Pinsonneault}, {Poovelil}, {Povick}, {Prakash}, {Price-Whelan}, {Raddick},
  {Raichoor}, {Ray}, {Rembold}, {Rezaie}, {Riffel}, {Riffel}, {Rix}, {Robin},
  {Roman-Lopes}, {Rom{\'a}n-Z{\'u}{\~n}iga}, {Rose}, {Ross}, {Rossi},
  {Rowlands}, {Rubin}, {Salvato}, {S{\'a}nchez}, {S{\'a}nchez-Menguiano},
  {S{\'a}nchez-Gallego}, {Sayres}, {Schaefer}, {Schiavon}, {Schimoia},
  {Schlafly}, {Schlegel}, {Schneider}, {Schultheis}, {Schwope}, {Seo},
  {Serenelli}, {Shafieloo}, {Shamsi}, {Shao}, {Shen}, {Shetrone}, {Shirley},
  {Aguirre}, {Simon}, {Skrutskie}, {Slosar}, {Smethurst}, {Sobeck}, {Sodi},
  {Souto}, {Stark}, {Stassun}, {Steinmetz}, {Stello}, {Stermer},
  {Storchi-Bergmann}, {Streblyanska}, {Stringfellow}, {Stutz}, {Su{\'a}rez},
  {Sun}, {Taghizadeh-Popp}, {Talbot}, {Tayar}, {Thakar}, {Theriault}, {Thomas},
  {Thomas}, {Tinker}, {Tojeiro}, {Toledo}, {Tremonti}, {Troup}, {Tuttle},
  {Unda-Sanzana}, {Valentini}, {Vargas-Gonz{\'a}lez}, {Vargas-Maga{\~n}a},
  {V{\'a}zquez-Mata}, {Vivek}, {Wake}, {Wang}, {Weaver}, {Weijmans}, {Wild},
  {Wilson}, {Wilson}, {Wolthuis}, {Wood-Vasey}, {Yan}, {Yang}, {Y{\`e}che},
  {Zamora}, {Zarrouk}, {Zasowski}, {Zhang}, {Zhao}, {Zhao}, {Zheng}, {Zheng},
  {Zhu}, \& {Zou}}]{Ahumada20}
{Ahumada}, R., {Prieto}, C.~A., {Almeida}, A., {et~al.} 2020, \apjs, 249, 3,
  \dodoi{10.3847/1538-4365/ab929e}

\bibitem[{{Aihara} {et~al.}(2022){Aihara}, {AlSayyad}, {Ando}, {Armstrong},
  {Bosch}, {Egami}, {Furusawa}, {Furusawa}, {Harasawa}, {Harikane}, {Hsieh},
  {Ikeda}, {Ito}, {Iwata}, {Kodama}, {Koike}, {Kokubo}, {Komiyama}, {Li},
  {Liang}, {Lin}, {Lupton}, {Lust}, {MacArthur}, {Mawatari}, {Mineo},
  {Miyatake}, {Miyazaki}, {More}, {Morishima}, {Murayama}, {Nakajima},
  {Nakata}, {Nishizawa}, {Oguri}, {Okabe}, {Okura}, {Ono}, {Osato}, {Ouchi},
  {Pan}, {Plazas Malag{\'o}n}, {Price}, {Reed}, {Rykoff}, {Shibuya},
  {Simunovic}, {Strauss}, {Sugimori}, {Suto}, {Suzuki}, {Takada}, {Takagi},
  {Takata}, {Takita}, {Tanaka}, {Tang}, {Taranu}, {Terai}, {Toba}, {Turner},
  {Uchiyama}, {Vijarnwannaluk}, {Waters}, {Yamada}, {Yamamoto}, \&
  {Yamashita}}]{Aihara22}
{Aihara}, H., {AlSayyad}, Y., {Ando}, M., {et~al.} 2022, \pasj, 74, 247,
  \dodoi{10.1093/pasj/psab122}

\bibitem[{{Amiaux} {et~al.}(2012){Amiaux}, {Scaramella}, {Mellier}, {Altieri},
  {Burigana}, {Da Silva}, {Gomez}, {Hoar}, {Laureijs}, {Maiorano},
  {Magalh{\~a}es Oliveira}, {Renk}, {Saavedra Criado}, {Tereno},
  {Augu{\`e}res}, {Brinchmann}, {Cropper}, {Duvet}, {Ealet}, {Franzetti},
  {Garilli}, {Gondoin}, {Guzzo}, {Hoekstra}, {Holmes}, {Jahnke}, {Kitching},
  {Meneghetti}, {Percival}, \& {Warren}}]{Amiaux12}
{Amiaux}, J., {Scaramella}, R., {Mellier}, Y., {et~al.} 2012, in Society of
  Photo-Optical Instrumentation Engineers (SPIE) Conference Series, Vol. 8442,
  Space Telescopes and Instrumentation 2012: Optical, Infrared, and Millimeter
  Wave, ed. M.~C. {Clampin}, G.~G. {Fazio}, H.~A. {MacEwen}, \& J.~{Oschmann},
  Jacobus~M., 84420Z, \dodoi{10.1117/12.926513}

\bibitem[{{Arnouts} {et~al.}(1999){Arnouts}, {Cristiani}, {Moscardini},
  {Matarrese}, {Lucchin}, {Fontana}, \& {Giallongo}}]{Arnouts99}
{Arnouts}, S., {Cristiani}, S., {Moscardini}, L., {et~al.} 1999, \mnras, 310,
  540, \dodoi{10.1046/j.1365-8711.1999.02978.x}

\bibitem[{{Astropy Collaboration} {et~al.}(2013){Astropy Collaboration},
  {Robitaille}, {Tollerud}, {Greenfield}, {Droettboom}, {Bray}, {Aldcroft},
  {Davis}, {Ginsburg}, {Price-Whelan}, {Kerzendorf}, {Conley}, {Crighton},
  {Barbary}, {Muna}, {Ferguson}, {Grollier}, {Parikh}, {Nair}, {Unther},
  {Deil}, {Woillez}, {Conseil}, {Kramer}, {Turner}, {Singer}, {Fox}, {Weaver},
  {Zabalza}, {Edwards}, {Azalee Bostroem}, {Burke}, {Casey}, {Crawford},
  {Dencheva}, {Ely}, {Jenness}, {Labrie}, {Lim}, {Pierfederici}, {Pontzen},
  {Ptak}, {Refsdal}, {Servillat}, \& {Streicher}}]{Astropy13}
{Astropy Collaboration}, {Robitaille}, T.~P., {Tollerud}, E.~J., {et~al.} 2013,
  \aap, 558, A33, \dodoi{10.1051/0004-6361/201322068}

\bibitem[{{Astropy Collaboration} {et~al.}(2018){Astropy Collaboration},
  {Price-Whelan}, {Sip{\H{o}}cz}, {G{\"u}nther}, {Lim}, {Crawford}, {Conseil},
  {Shupe}, {Craig}, {Dencheva}, {Ginsburg}, {VanderPlas}, {Bradley},
  {P{\'e}rez-Su{\'a}rez}, {de Val-Borro}, {Aldcroft}, {Cruz}, {Robitaille},
  {Tollerud}, {Ardelean}, {Babej}, {Bach}, {Bachetti}, {Bakanov}, {Bamford},
  {Barentsen}, {Barmby}, {Baumbach}, {Berry}, {Biscani}, {Boquien}, {Bostroem},
  {Bouma}, {Brammer}, {Bray}, {Breytenbach}, {Buddelmeijer}, {Burke},
  {Calderone}, {Cano Rodr{\'\i}guez}, {Cara}, {Cardoso}, {Cheedella}, {Copin},
  {Corrales}, {Crichton}, {D'Avella}, {Deil}, {Depagne}, {Dietrich}, {Donath},
  {Droettboom}, {Earl}, {Erben}, {Fabbro}, {Ferreira}, {Finethy}, {Fox},
  {Garrison}, {Gibbons}, {Goldstein}, {Gommers}, {Greco}, {Greenfield},
  {Groener}, {Grollier}, {Hagen}, {Hirst}, {Homeier}, {Horton}, {Hosseinzadeh},
  {Hu}, {Hunkeler}, {Ivezi{\'c}}, {Jain}, {Jenness}, {Kanarek}, {Kendrew},
  {Kern}, {Kerzendorf}, {Khvalko}, {King}, {Kirkby}, {Kulkarni}, {Kumar},
  {Lee}, {Lenz}, {Littlefair}, {Ma}, {Macleod}, {Mastropietro}, {McCully},
  {Montagnac}, {Morris}, {Mueller}, {Mumford}, {Muna}, {Murphy}, {Nelson},
  {Nguyen}, {Ninan}, {N{\"o}the}, {Ogaz}, {Oh}, {Parejko}, {Parley}, {Pascual},
  {Patil}, {Patil}, {Plunkett}, {Prochaska}, {Rastogi}, {Reddy Janga},
  {Sabater}, {Sakurikar}, {Seifert}, {Sherbert}, {Sherwood-Taylor}, {Shih},
  {Sick}, {Silbiger}, {Singanamalla}, {Singer}, {Sladen}, {Sooley},
  {Sornarajah}, {Streicher}, {Teuben}, {Thomas}, {Tremblay}, {Turner},
  {Terr{\'o}n}, {van Kerkwijk}, {de la Vega}, {Watkins}, {Weaver}, {Whitmore},
  {Woillez}, {Zabalza}, \& {Astropy Contributors}}]{Astropy18}
{Astropy Collaboration}, {Price-Whelan}, A.~M., {Sip{\H{o}}cz}, B.~M., {et~al.}
  2018, \aj, 156, 123, \dodoi{10.3847/1538-3881/aabc4f}

\bibitem[{{Balogh} {et~al.}(1999){Balogh}, {Morris}, {Yee}, {Carlberg}, \&
  {Ellingson}}]{Balogh99}
{Balogh}, M.~L., {Morris}, S.~L., {Yee}, H.~K.~C., {Carlberg}, R.~G., \&
  {Ellingson}, E. 1999, \apj, 527, 54, \dodoi{10.1086/308056}

\bibitem[{{Barone} {et~al.}(2022){Barone}, {D'Eugenio}, {Scott}, {Colless},
  {Vaughan}, {van der Wel}, {Fraser-McKelvie}, {de Graaff}, {van de Sande},
  {Wu}, {Bezanson}, {Brough}, {Bell}, {Croom}, {Cortese}, {Driver}, {Gallazzi},
  {Muzzin}, {Sobral}, {Bland-Hawthorn}, {Bryant}, {Goodwin}, {Lawrence},
  {Lorente}, \& {Owers}}]{Barone22}
{Barone}, T.~M., {D'Eugenio}, F., {Scott}, N., {et~al.} 2022, \mnras, 512,
  3828, \dodoi{10.1093/mnras/stac705}

\bibitem[{{Beifiori} {et~al.}(2014){Beifiori}, {Thomas}, {Maraston}, {Steele},
  {Masters}, {Pforr}, {Saglia}, {Bender}, {Tojeiro}, {Chen}, {Bolton},
  {Brownstein}, {Johansson}, {Leauthaud}, {Nichol}, {Schneider}, {Senger},
  {Skibba}, {Wake}, {Pan}, {Snedden}, {Bizyaev}, {Brewington}, {Malanushenko},
  {Malanushenko}, {Oravetz}, {Simmons}, {Shelden}, \& {Ebelke}}]{Beifiori14}
{Beifiori}, A., {Thomas}, D., {Maraston}, C., {et~al.} 2014, \apj, 789, 92,
  \dodoi{10.1088/0004-637X/789/2/92}

\bibitem[{{Belli} {et~al.}(2015){Belli}, {Newman}, \& {Ellis}}]{Belli15}
{Belli}, S., {Newman}, A.~B., \& {Ellis}, R.~S. 2015, \apj, 799, 206,
  \dodoi{10.1088/0004-637X/799/2/206}

\bibitem[{{Bertin}(2011)}]{Bertin11}
{Bertin}, E. 2011, in Astronomical Society of the Pacific Conference Series,
  Vol. 442, Astronomical Data Analysis Software and Systems XX, ed. I.~N.
  {Evans}, A.~{Accomazzi}, D.~J. {Mink}, \& A.~H. {Rots}, 435

\bibitem[{{Bertin} \& {Arnouts}(1996)}]{Bertin96}
{Bertin}, E., \& {Arnouts}, S. 1996, \aaps, 117, 393,
  \dodoi{10.1051/aas:1996164}

\bibitem[{{Bezanson} {et~al.}(2009){Bezanson}, {van Dokkum}, {Tal},
  {Marchesini}, {Kriek}, {Franx}, \& {Coppi}}]{Bezanson09}
{Bezanson}, R., {van Dokkum}, P.~G., {Tal}, T., {et~al.} 2009, \apj, 697, 1290,
  \dodoi{10.1088/0004-637X/697/2/1290}

\bibitem[{{Bezanson} {et~al.}(2013){Bezanson}, {van Dokkum}, {van de Sande},
  {Franx}, {Leja}, \& {Kriek}}]{Bezanson13}
{Bezanson}, R., {van Dokkum}, P.~G., {van de Sande}, J., {et~al.} 2013, \apjl,
  779, L21, \dodoi{10.1088/2041-8205/779/2/L21}

\bibitem[{{Bois} {et~al.}(2011){Bois}, {Emsellem}, {Bournaud}, {Alatalo},
  {Blitz}, {Bureau}, {Cappellari}, {Davies}, {Davis}, {de Zeeuw}, {Duc},
  {Khochfar}, {Krajnovi{\'c}}, {Kuntschner}, {Lablanche}, {McDermid},
  {Morganti}, {Naab}, {Oosterloo}, {Sarzi}, {Scott}, {Serra}, {Weijmans}, \&
  {Young}}]{Bois11}
{Bois}, M., {Emsellem}, E., {Bournaud}, F., {et~al.} 2011, \mnras, 416, 1654,
  \dodoi{10.1111/j.1365-2966.2011.19113.x}

\bibitem[{{Bosch} {et~al.}(2018){Bosch}, {Armstrong}, {Bickerton}, {Furusawa},
  {Ikeda}, {Koike}, {Lupton}, {Mineo}, {Price}, {Takata}, {Tanaka}, {Yasuda},
  {AlSayyad}, {Becker}, {Coulton}, {Coupon}, {Garmilla}, {Huang}, {Krughoff},
  {Lang}, {Leauthaud}, {Lim}, {Lust}, {MacArthur}, {Mandelbaum}, {Miyatake},
  {Miyazaki}, {Murata}, {More}, {Okura}, {Owen}, {Swinbank}, {Strauss},
  {Yamada}, \& {Yamanoi}}]{Bosch18}
{Bosch}, J., {Armstrong}, R., {Bickerton}, S., {et~al.} 2018, \pasj, 70, S5,
  \dodoi{10.1093/pasj/psx080}

\bibitem[{{Bruzual} \& {Charlot}(2003)}]{Bruzual03}
{Bruzual}, G., \& {Charlot}, S. 2003, \mnras, 344, 1000,
  \dodoi{10.1046/j.1365-8711.2003.06897.x}

\bibitem[{{Buitrago} {et~al.}(2017){Buitrago}, {Trujillo}, {Curtis-Lake},
  {Montes}, {Cooper}, {Bruce}, {P{\'e}rez-Gonz{\'a}lez}, \&
  {Cirasuolo}}]{Buitrago17}
{Buitrago}, F., {Trujillo}, I., {Curtis-Lake}, E., {et~al.} 2017, \mnras, 466,
  4888, \dodoi{10.1093/mnras/stw3382}

\bibitem[{{Calzetti} {et~al.}(2000){Calzetti}, {Armus}, {Bohlin}, {Kinney},
  {Koornneef}, \& {Storchi-Bergmann}}]{Calzetti00}
{Calzetti}, D., {Armus}, L., {Bohlin}, R.~C., {et~al.} 2000, \apj, 533, 682,
  \dodoi{10.1086/308692}

\bibitem[{{Carollo} {et~al.}(2013){Carollo}, {Bschorr}, {Renzini}, {Lilly},
  {Capak}, {Cibinel}, {Ilbert}, {Onodera}, {Scoville}, {Cameron}, {Mobasher},
  {Sanders}, \& {Taniguchi}}]{Carollo13}
{Carollo}, C.~M., {Bschorr}, T.~J., {Renzini}, A., {et~al.} 2013, \apj, 773,
  112, \dodoi{10.1088/0004-637X/773/2/112}

\bibitem[{{Cassata} {et~al.}(2011){Cassata}, {Giavalisco}, {Guo}, {Renzini},
  {Ferguson}, {Koekemoer}, {Salimbeni}, {Scarlata}, {Grogin}, {Conselice},
  {Dahlen}, {Lotz}, {Dickinson}, \& {Lin}}]{Cassata11}
{Cassata}, P., {Giavalisco}, M., {Guo}, Y., {et~al.} 2011, \apj, 743, 96,
  \dodoi{10.1088/0004-637X/743/1/96}

\bibitem[{{Cassata} {et~al.}(2013){Cassata}, {Giavalisco}, {Williams}, {Guo},
  {Lee}, {Renzini}, {Ferguson}, {Faber}, {Barro}, {McIntosh}, {Lu}, {Bell},
  {Koo}, {Papovich}, {Ryan}, {Conselice}, {Grogin}, {Koekemoer}, \&
  {Hathi}}]{Cassata13}
{Cassata}, P., {Giavalisco}, M., {Williams}, C.~C., {et~al.} 2013, \apj, 775,
  106, \dodoi{10.1088/0004-637X/775/2/106}

\bibitem[{{Chabrier}(2003)}]{Chabrier03}
{Chabrier}, G. 2003, \pasp, 115, 763, \dodoi{10.1086/376392}

\bibitem[{{Chan} {et~al.}(2016){Chan}, {Beifiori}, {Mendel}, {Saglia},
  {Bender}, {Fossati}, {Galametz}, {Wegner}, {Wilman}, {Cappellari}, {Davies},
  {Houghton}, {Prichard}, {Lewis}, {Sharples}, \& {Stott}}]{Chan16}
{Chan}, J. C.~C., {Beifiori}, A., {Mendel}, J.~T., {et~al.} 2016, \mnras, 458,
  3181, \dodoi{10.1093/mnras/stw502}

\bibitem[{{Charbonnier} {et~al.}(2017){Charbonnier}, {Huertas-Company},
  {Gon{\c{c}}alves}, {Men{\'e}ndez-Delmestre}, {Bundy}, {Galliano}, {Moraes},
  {Makler}, {Pereira}, {Erben}, {Hildebrandt}, {Shan}, {Caminha}, {Grossi}, \&
  {Riguccini}}]{Charbonnier17}
{Charbonnier}, A., {Huertas-Company}, M., {Gon{\c{c}}alves}, T.~S., {et~al.}
  2017, \mnras, 469, 4523, \dodoi{10.1093/mnras/stx1142}

\bibitem[{{Cimatti} {et~al.}(2012){Cimatti}, {Nipoti}, \&
  {Cassata}}]{Cimatti12}
{Cimatti}, A., {Nipoti}, C., \& {Cassata}, P. 2012, \mnras, 422, L62,
  \dodoi{10.1111/j.1745-3933.2012.01237.x}

\bibitem[{{Conroy} \& {Gunn}(2010)}]{Conroy2010}
{Conroy}, C., \& {Gunn}, J.~E. 2010, {FSPS: Flexible Stellar Population
  Synthesis}, Astrophysics Source Code Library, record ascl:1010.043.
\newblock \doeprint{1010.043}

\bibitem[{{Conroy} {et~al.}(2009){Conroy}, {Gunn}, \& {White}}]{conroy2009}
{Conroy}, C., {Gunn}, J.~E., \& {White}, M. 2009, \apj, 699, 486,
  \dodoi{10.1088/0004-637X/699/1/486}

\bibitem[{{Conselice}(2014)}]{Conselice14}
{Conselice}, C.~J. 2014, \araa, 52, 291,
  \dodoi{10.1146/annurev-astro-081913-040037}

\bibitem[{{Cowie} {et~al.}(1996){Cowie}, {Songaila}, {Hu}, \&
  {Cohen}}]{Cowie96}
{Cowie}, L.~L., {Songaila}, A., {Hu}, E.~M., \& {Cohen}, J.~G. 1996, \aj, 112,
  839, \dodoi{10.1086/118058}

\bibitem[{{Daddi} {et~al.}(2005){Daddi}, {Renzini}, {Pirzkal}, {Cimatti},
  {Malhotra}, {Stiavelli}, {Xu}, {Pasquali}, {Rhoads}, {Brusa}, {di Serego
  Alighieri}, {Ferguson}, {Koekemoer}, {Moustakas}, {Panagia}, \&
  {Windhorst}}]{Daddi05}
{Daddi}, E., {Renzini}, A., {Pirzkal}, N., {et~al.} 2005, \apj, 626, 680,
  \dodoi{10.1086/430104}

\bibitem[{{Damjanov} {et~al.}(2022){Damjanov}, {Sohn}, {Utsumi}, {Geller}, \&
  {Dell'Antonio}}]{Damjanov22}
{Damjanov}, I., {Sohn}, J., {Utsumi}, Y., {Geller}, M.~J., \& {Dell'Antonio},
  I. 2022, \apj, 929, 61, \dodoi{10.3847/1538-4357/ac54bd}

\bibitem[{{Damjanov} {et~al.}(2018){Damjanov}, {Zahid}, {Geller}, {Fabricant},
  \& {Hwang}}]{Damjanov18}
{Damjanov}, I., {Zahid}, H.~J., {Geller}, M.~J., {Fabricant}, D.~G., \&
  {Hwang}, H.~S. 2018, \apjs, 234, 21, \dodoi{10.3847/1538-4365/aaa01c}

\bibitem[{{Damjanov} {et~al.}(2019){Damjanov}, {Zahid}, {Geller}, {Utsumi},
  {Sohn}, \& {Souchereau}}]{Damjanov19}
{Damjanov}, I., {Zahid}, H.~J., {Geller}, M.~J., {et~al.} 2019, \apj, 872, 91,
  \dodoi{10.3847/1538-4357/aaf97d}

\bibitem[{{Damjanov} {et~al.}(2011){Damjanov}, {Abraham}, {Glazebrook},
  {McCarthy}, {Caris}, {Carlberg}, {Chen}, {Crampton}, {Green}, {J{\o}rgensen},
  {Juneau}, {Le Borgne}, {Marzke}, {Mentuch}, {Murowinski}, {Roth}, {Savaglio},
  \& {Yan}}]{Damjanov11}
{Damjanov}, I., {Abraham}, R.~G., {Glazebrook}, K., {et~al.} 2011, \apjl, 739,
  L44, \dodoi{10.1088/2041-8205/739/2/L44}

\bibitem[{{de Graaff} {et~al.}(2022){de Graaff}, {Trayford}, {Franx},
  {Schaller}, {Schaye}, \& {van der Wel}}]{deGraff22}
{de Graaff}, A., {Trayford}, J., {Franx}, M., {et~al.} 2022, \mnras, 511, 2544,
  \dodoi{10.1093/mnras/stab3510}

\bibitem[{{D{\'\i}az-Garc{\'\i}a} {et~al.}(2019){D{\'\i}az-Garc{\'\i}a},
  {Cenarro}, {L{\'o}pez-Sanjuan}, {Ferreras}, {Fern{\'a}ndez-Soto},
  {Gonz{\'a}lez Delgado}, {M{\'a}rquez}, {Masegosa}, {San Roman}, {Viironen},
  {Bonoli}, {Cervi{\~n}o}, {Moles}, {Crist{\'o}bal-Hornillos}, {Alfaro},
  {Aparicio-Villegas}, {Ben{\'\i}tez}, {Broadhurst}, {Cabrera-Ca{\~n}o},
  {Castander}, {Cepa}, {Husillos}, {Infante}, {Aguerri}, {Mart{\'\i}nez},
  {Molino}, {del Olmo}, {Perea}, {Prada}, \& {Quintana}}]{DiazGarcia19}
{D{\'\i}az-Garc{\'\i}a}, L.~A., {Cenarro}, A.~J., {L{\'o}pez-Sanjuan}, C.,
  {et~al.} 2019, \aap, 631, A157, \dodoi{10.1051/0004-6361/201832882}

\bibitem[{{Driver} \& {Robotham}(2010)}]{Driver10}
{Driver}, S.~P., \& {Robotham}, A. S.~G. 2010, \mnras, 407, 2131,
  \dodoi{10.1111/j.1365-2966.2010.17028.x}

\bibitem[{{Fabricant} {et~al.}(2005){Fabricant}, {Fata}, {Roll}, {Hertz},
  {Caldwell}, {Gauron}, {Geary}, {McLeod}, {Szentgyorgyi}, {Zajac}, {Kurtz},
  {Barberis}, {Bergner}, {Brown}, {Conroy}, {Eng}, {Geller}, {Goddard},
  {Honsa}, {Mueller}, {Mink}, {Ordway}, {Tokarz}, {Woods}, {Wyatt}, {Epps}, \&
  {Dell'Antonio}}]{Fabricant05}
{Fabricant}, D., {Fata}, R., {Roll}, J., {et~al.} 2005, \pasp, 117, 1411,
  \dodoi{10.1086/497385}

\bibitem[{{Fagioli} {et~al.}(2016){Fagioli}, {Carollo}, {Renzini}, {Lilly},
  {Onodera}, \& {Tacchella}}]{Fagioli16}
{Fagioli}, M., {Carollo}, C.~M., {Renzini}, A., {et~al.} 2016, \apj, 831, 173,
  \dodoi{10.3847/0004-637X/831/2/173}

\bibitem[{{Faisst} {et~al.}(2017){Faisst}, {Carollo}, {Capak}, {Tacchella},
  {Renzini}, {Ilbert}, {McCracken}, \& {Scoville}}]{Faisst17}
{Faisst}, A.~L., {Carollo}, C.~M., {Capak}, P.~L., {et~al.} 2017, \apj, 839,
  71, \dodoi{10.3847/1538-4357/aa697a}

\bibitem[{{Fan} {et~al.}(2010){Fan}, {Lapi}, {Bressan}, {Bernardi}, {De Zotti},
  \& {Danese}}]{Fan10}
{Fan}, L., {Lapi}, A., {Bressan}, A., {et~al.} 2010, \apj, 718, 1460,
  \dodoi{10.1088/0004-637X/718/2/1460}

\bibitem[{{Fan} {et~al.}(2008){Fan}, {Lapi}, {De Zotti}, \& {Danese}}]{Fan08}
{Fan}, L., {Lapi}, A., {De Zotti}, G., \& {Danese}, L. 2008, \apjl, 689, L101,
  \dodoi{10.1086/595784}

\bibitem[{{Geller} \& {Hwang}(2015)}]{Geller15}
{Geller}, M.~J., \& {Hwang}, H.~S. 2015, Astronomische Nachrichten, 336, 428,
  \dodoi{10.1002/asna.201512182}

\bibitem[{{Geller} {et~al.}(2016){Geller}, {Hwang}, {Dell'Antonio}, {Zahid},
  {Kurtz}, \& {Fabricant}}]{Geller16}
{Geller}, M.~J., {Hwang}, H.~S., {Dell'Antonio}, I.~P., {et~al.} 2016, \apjs,
  224, 11, \dodoi{10.3847/0067-0049/224/1/11}

\bibitem[{{Geller} {et~al.}(2014){Geller}, {Hwang}, {Fabricant}, {Kurtz},
  {Dell'Antonio}, \& {Zahid}}]{Geller14}
{Geller}, M.~J., {Hwang}, H.~S., {Fabricant}, D.~G., {et~al.} 2014, \apjs, 213,
  35, \dodoi{10.1088/0067-0049/213/2/35}

\bibitem[{{Genel} {et~al.}(2018){Genel}, {Nelson}, {Pillepich}, {Springel},
  {Pakmor}, {Weinberger}, {Hernquist}, {Naiman}, {Vogelsberger}, {Marinacci},
  \& {Torrey}}]{Genel18}
{Genel}, S., {Nelson}, D., {Pillepich}, A., {et~al.} 2018, \mnras, 474, 3976,
  \dodoi{10.1093/mnras/stx3078}

\bibitem[{{Greene} {et~al.}(2022){Greene}, {Bezanson}, {Ouchi}, {Silverman}, \&
  {the PFS Galaxy Evolution Working Group}}]{Greene22}
{Greene}, J., {Bezanson}, R., {Ouchi}, M., {Silverman}, J., \& {the PFS Galaxy
  Evolution Working Group}. 2022, arXiv e-prints, arXiv:2206.14908.
\newblock \doarXiv{2206.14908}

\bibitem[{{Greene} {et~al.}(2015){Greene}, {Janish}, {Ma}, {McConnell},
  {Blakeslee}, {Thomas}, \& {Murphy}}]{Greene15}
{Greene}, J.~E., {Janish}, R., {Ma}, C.-P., {et~al.} 2015, \apj, 807, 11,
  \dodoi{10.1088/0004-637X/807/1/11}

\bibitem[{{Haines} {et~al.}(2017){Haines}, {Iovino}, {Krywult}, {Guzzo},
  {Davidzon}, {Bolzonella}, {Garilli}, {Scodeggio}, {Granett}, {de la Torre},
  {De Lucia}, {Abbas}, {Adami}, {Arnouts}, {Bottini}, {Cappi}, {Cucciati},
  {Franzetti}, {Fritz}, {Gargiulo}, {Le Brun}, {Le F{\`e}vre}, {Maccagni},
  {Ma{\l}ek}, {Marulli}, {Moutard}, {Polletta}, {Pollo}, {Tasca}, {Tojeiro},
  {Vergani}, {Zanichelli}, {Zamorani}, {Bel}, {Branchini}, {Coupon}, {Ilbert},
  {Moscardini}, {Peacock}, \& {Siudek}}]{Haines17}
{Haines}, C.~P., {Iovino}, A., {Krywult}, J., {et~al.} 2017, \aap, 605, A4,
  \dodoi{10.1051/0004-6361/201630118}

\bibitem[{{Hamadouche} {et~al.}(2022){Hamadouche}, {Carnall}, {McLure},
  {Dunlop}, {McLeod}, {Cullen}, {Begley}, {Bolzonella}, {Buitrago},
  {Castellano}, {Cucciati}, {Fontana}, {Gargiulo}, {Moresco}, {Pozzetti}, \&
  {Zamorani}}]{Hamadouche22}
{Hamadouche}, M.~L., {Carnall}, A.~C., {McLure}, R.~J., {et~al.} 2022, \mnras,
  512, 1262, \dodoi{10.1093/mnras/stac535}

\bibitem[{Harris {et~al.}(2020)Harris, Millman, van~der Walt, Gommers,
  Virtanen, Cournapeau, Wieser, Taylor, Berg, Smith, Kern, Picus, Hoyer, van
  Kerkwijk, Brett, Haldane, Fernández~del Río, Wiebe, Peterson,
  Gérard-Marchant, Sheppard, Reddy, Weckesser, Abbasi, Gohlke, \&
  Oliphant}]{2020NumPy-Array}
Harris, C.~R., Millman, K.~J., van~der Walt, S.~J., {et~al.} 2020, Nature, 585,
  357–362, \dodoi{10.1038/s41586-020-2649-2}

\bibitem[{{Hill} {et~al.}(2017){Hill}, {Muzzin}, {Franx}, {Clauwens},
  {Schreiber}, {Marchesini}, {Stefanon}, {Labbe}, {Brammer}, {Caputi}, {Fynbo},
  {Milvang-Jensen}, {Skelton}, {van Dokkum}, \& {Whitaker}}]{Hill17}
{Hill}, A.~R., {Muzzin}, A., {Franx}, M., {et~al.} 2017, \apj, 837, 147,
  \dodoi{10.3847/1538-4357/aa61fe}

\bibitem[{{Hogg}(1999)}]{Hogg99}
{Hogg}, D.~W. 1999, arXiv e-prints, astro.
\newblock \doarXiv{astro-ph/9905116}

\bibitem[{{Hopkins} {et~al.}(2010){Hopkins}, {Bundy}, {Hernquist}, {Wuyts}, \&
  {Cox}}]{Hopkins10a}
{Hopkins}, P.~F., {Bundy}, K., {Hernquist}, L., {Wuyts}, S., \& {Cox}, T.~J.
  2010, \mnras, 401, 1099, \dodoi{10.1111/j.1365-2966.2009.15699.x}

\bibitem[{{Hopkins} {et~al.}(2009{\natexlab{a}}){Hopkins}, {Bundy}, {Murray},
  {Quataert}, {Lauer}, \& {Ma}}]{Hopkins09a}
{Hopkins}, P.~F., {Bundy}, K., {Murray}, N., {et~al.} 2009{\natexlab{a}},
  \mnras, 398, 898, \dodoi{10.1111/j.1365-2966.2009.15062.x}

\bibitem[{{Hopkins} {et~al.}(2009{\natexlab{b}}){Hopkins}, {Hernquist}, {Cox},
  {Keres}, \& {Wuyts}}]{Hopkins09b}
{Hopkins}, P.~F., {Hernquist}, L., {Cox}, T.~J., {Keres}, D., \& {Wuyts}, S.
  2009{\natexlab{b}}, \apj, 691, 1424, \dodoi{10.1088/0004-637X/691/2/1424}

\bibitem[{{Huertas-Company} {et~al.}(2013){Huertas-Company}, {Mei}, {Shankar},
  {Delaye}, {Raichoor}, {Covone}, {Finoguenov}, {Kneib}, {Le}, \&
  {Povic}}]{Huertas-Company13}
{Huertas-Company}, M., {Mei}, S., {Shankar}, F., {et~al.} 2013, \mnras, 428,
  1715, \dodoi{10.1093/mnras/sts150}

\bibitem[{{Huertas-Company} {et~al.}(2015){Huertas-Company},
  {P{\'e}rez-Gonz{\'a}lez}, {Mei}, {Shankar}, {Bernardi}, {Daddi}, {Barro},
  {Cabrera-Vives}, {Cattaneo}, {Dimauro}, \& {Gravet}}]{Huertas-Company15}
{Huertas-Company}, M., {P{\'e}rez-Gonz{\'a}lez}, P.~G., {Mei}, S., {et~al.}
  2015, \apj, 809, 95, \dodoi{10.1088/0004-637X/809/1/95}

\bibitem[{{Hyde} \& {Bernardi}(2009)}]{Hyde09}
{Hyde}, J.~B., \& {Bernardi}, M. 2009, \mnras, 396, 1171,
  \dodoi{10.1111/j.1365-2966.2009.14783.x}

\bibitem[{{Ilbert} {et~al.}(2006){Ilbert}, {Arnouts}, {McCracken},
  {Bolzonella}, {Bertin}, {Le F{\`e}vre}, {Mellier}, {Zamorani}, {Pell{\`o}},
  {Iovino}, {Tresse}, {Le Brun}, {Bottini}, {Garilli}, {Maccagni}, {Picat},
  {Scaramella}, {Scodeggio}, {Vettolani}, {Zanichelli}, {Adami}, {Bardelli},
  {Cappi}, {Charlot}, {Ciliegi}, {Contini}, {Cucciati}, {Foucaud}, {Franzetti},
  {Gavignaud}, {Guzzo}, {Marano}, {Marinoni}, {Mazure}, {Meneux}, {Merighi},
  {Paltani}, {Pollo}, {Pozzetti}, {Radovich}, {Zucca}, {Bondi}, {Bongiorno},
  {Busarello}, {de La Torre}, {Gregorini}, {Lamareille}, {Mathez}, {Merluzzi},
  {Ripepi}, {Rizzo}, \& {Vergani}}]{Ilbert06}
{Ilbert}, O., {Arnouts}, S., {McCracken}, H.~J., {et~al.} 2006, \aap, 457, 841,
  \dodoi{10.1051/0004-6361:20065138}

\bibitem[{Jekel \& Venter(2019)}]{pwlf}
Jekel, C.~F., \& Venter, G. 2019, {pwlf:} A Python Library for Fitting 1D
  Continuous Piecewise Linear Functions.
\newblock \url{https://github.com/cjekel/piecewise_linear_fit_py}

\bibitem[{{Kauffmann} {et~al.}(2003){Kauffmann}, {Heckman}, {White}, {Charlot},
  {Tremonti}, {Brinchmann}, {Bruzual}, {Peng}, {Seibert}, {Bernardi},
  {Blanton}, {Brinkmann}, {Castander}, {Cs{\'a}bai}, {Fukugita}, {Ivezic},
  {Munn}, {Nichol}, {Padmanabhan}, {Thakar}, {Weinberg}, \&
  {York}}]{Kauffmann03}
{Kauffmann}, G., {Heckman}, T.~M., {White}, S. D.~M., {et~al.} 2003, \mnras,
  341, 33, \dodoi{10.1046/j.1365-8711.2003.06291.x}

\bibitem[{{Kawinwanichakij} {et~al.}(2021){Kawinwanichakij}, {Silverman},
  {Ding}, {George}, {Damjanov}, {Sawicki}, {Tanaka}, {Taranu}, {Birrer},
  {Huang}, {Li}, {Onodera}, {Shibuya}, \& {Yasuda}}]{Kawin21}
{Kawinwanichakij}, L., {Silverman}, J.~D., {Ding}, X., {et~al.} 2021, \apj,
  921, 38, \dodoi{10.3847/1538-4357/ac1f21}

\bibitem[{{L{\'o}pez-Sanjuan} {et~al.}(2012){L{\'o}pez-Sanjuan}, {Le
  F{\`e}vre}, {Ilbert}, {Tasca}, {Bridge}, {Cucciati}, {Kampczyk}, {Pozzetti},
  {Xu}, {Carollo}, {Contini}, {Kneib}, {Lilly}, {Mainieri}, {Renzini},
  {Sanders}, {Scodeggio}, {Scoville}, {Taniguchi}, {Zamorani}, {Aussel},
  {Bardelli}, {Bolzonella}, {Bongiorno}, {Capak}, {Caputi}, {de la Torre}, {de
  Ravel}, {Franzetti}, {Garilli}, {Iovino}, {Knobel}, {Kova{\v{c}}},
  {Lamareille}, {Le Borgne}, {Le Brun}, {Le Floc'h}, {Maier}, {McCracken},
  {Mignoli}, {Pell{\'o}}, {Peng}, {P{\'e}rez-Montero}, {Presotto},
  {Ricciardelli}, {Salvato}, {Silverman}, {Tanaka}, {Tresse}, {Vergani},
  {Zucca}, {Barnes}, {Bordoloi}, {Cappi}, {Cimatti}, {Coppa}, {Koekemoer},
  {Liu}, {Moresco}, {Nair}, {Oesch}, {Schawinski}, \&
  {Welikala}}]{Lopez-Sanjuan12}
{L{\'o}pez-Sanjuan}, C., {Le F{\`e}vre}, O., {Ilbert}, O., {et~al.} 2012, \aap,
  548, A7, \dodoi{10.1051/0004-6361/201219085}

\bibitem[{{Lotz} {et~al.}(2011){Lotz}, {Jonsson}, {Cox}, {Croton}, {Primack},
  {Somerville}, \& {Stewart}}]{Lotz11}
{Lotz}, J.~M., {Jonsson}, P., {Cox}, T.~J., {et~al.} 2011, \apj, 742, 103,
  \dodoi{10.1088/0004-637X/742/2/103}

\bibitem[{{Mantha} {et~al.}(2018){Mantha}, {McIntosh}, {Brennan}, {Ferguson},
  {Kodra}, {Newman}, {Rafelski}, {Somerville}, {Conselice}, {Cook}, {Hathi},
  {Koo}, {Lotz}, {Simmons}, {Straughn}, {Snyder}, {Wuyts}, {Bell}, {Dekel},
  {Kartaltepe}, {Kocevski}, {Koekemoer}, {Lee}, {Lucas}, {Pacifici}, {Peth},
  {Barro}, {Dahlen}, {Finkelstein}, {Fontana}, {Galametz}, {Grogin}, {Guo},
  {Mobasher}, {Nayyeri}, {P{\'e}rez-Gonz{\'a}lez}, {Pforr}, {Santini},
  {Stefanon}, \& {Wiklind}}]{Mantha18}
{Mantha}, K.~B., {McIntosh}, D.~H., {Brennan}, R., {et~al.} 2018, \mnras, 475,
  1549, \dodoi{10.1093/mnras/stx3260}

\bibitem[{{Matharu} {et~al.}(2019){Matharu}, {Muzzin}, {Brammer}, {van der
  Burg}, {Auger}, {Hewett}, {van der Wel}, {van Dokkum}, {Balogh}, {Chan},
  {Demarco}, {Marchesini}, {Nelson}, {Noble}, {Wilson}, \& {Yee}}]{Matharu19}
{Matharu}, J., {Muzzin}, A., {Brammer}, G.~B., {et~al.} 2019, \mnras, 484, 595,
  \dodoi{10.1093/mnras/sty3465}

\bibitem[{{McLure} {et~al.}(2013){McLure}, {Pearce}, {Dunlop}, {Cirasuolo},
  {Curtis-Lake}, {Bruce}, {Caputi}, {Almaini}, {Bonfield}, {Bradshaw},
  {Buitrago}, {Chuter}, {Foucaud}, {Hartley}, \& {Jarvis}}]{McLure13}
{McLure}, R.~J., {Pearce}, H.~J., {Dunlop}, J.~S., {et~al.} 2013, \mnras, 428,
  1088, \dodoi{10.1093/mnras/sts092}

\bibitem[{{Miller} {et~al.}(2022){Miller}, {van Dokkum}, \& {Mowla}}]{Miller22}
{Miller}, T.~B., {van Dokkum}, P., \& {Mowla}, L. 2022, arXiv e-prints,
  arXiv:2207.05895.
\newblock \doarXiv{2207.05895}

\bibitem[{{Miyazaki} {et~al.}(2018){Miyazaki}, {Komiyama}, {Kawanomoto}, {Doi},
  {Furusawa}, {Hamana}, {Hayashi}, {Ikeda}, {Kamata}, {Karoji}, {Koike},
  {Kurakami}, {Miyama}, {Morokuma}, {Nakata}, {Namikawa}, {Nakaya}, {Nariai},
  {Obuchi}, {Oishi}, {Okada}, {Okura}, {Tait}, {Takata}, {Tanaka}, {Tanaka},
  {Terai}, {Tomono}, {Uraguchi}, {Usuda}, {Utsumi}, {Yamada}, {Yamanoi},
  {Aihara}, {Fujimori}, {Mineo}, {Miyatake}, {Oguri}, {Uchida}, {Tanaka},
  {Yasuda}, {Takada}, {Murayama}, {Nishizawa}, {Sugiyama}, {Chiba}, {Futamase},
  {Wang}, {Chen}, {Ho}, {Liaw}, {Chiu}, {Ho}, {Lai}, {Lee}, {Jeng}, {Iwamura},
  {Armstrong}, {Bickerton}, {Bosch}, {Gunn}, {Lupton}, {Loomis}, {Price},
  {Smith}, {Strauss}, {Turner}, {Suzuki}, {Miyazaki}, {Muramatsu}, {Yamamoto},
  {Endo}, {Ezaki}, {Ito}, {Kawaguchi}, {Sofuku}, {Taniike}, {Akutsu}, {Dojo},
  {Kasumi}, {Matsuda}, {Imoto}, {Miwa}, {Suzuki}, {Takeshi}, \&
  {Yokota}}]{Miyazaki18}
{Miyazaki}, S., {Komiyama}, Y., {Kawanomoto}, S., {et~al.} 2018, \pasj, 70, S1,
  \dodoi{10.1093/pasj/psx063}

\bibitem[{{Mosleh} {et~al.}(2020){Mosleh}, {Hosseinnejad},
  {Hosseini-ShahiSavandi}, \& {Tacchella}}]{Mosleh20}
{Mosleh}, M., {Hosseinnejad}, S., {Hosseini-ShahiSavandi}, S.~Z., \&
  {Tacchella}, S. 2020, \apj, 905, 170, \dodoi{10.3847/1538-4357/abc7cc}

\bibitem[{{Moustakas} {et~al.}(2013){Moustakas}, {Coil}, {Aird}, {Blanton},
  {Cool}, {Eisenstein}, {Mendez}, {Wong}, {Zhu}, \& {Arnouts}}]{Moustakas13}
{Moustakas}, J., {Coil}, A.~L., {Aird}, J., {et~al.} 2013, \apj, 767, 50,
  \dodoi{10.1088/0004-637X/767/1/50}

\bibitem[{{Mowla} {et~al.}(2019){Mowla}, {van Dokkum}, {Brammer}, {Momcheva},
  {van der Wel}, {Whitaker}, {Nelson}, {Bezanson}, {Muzzin}, {Franx},
  {MacKenty}, {Leja}, {Kriek}, \& {Marchesini}}]{Mowla19}
{Mowla}, L.~A., {van Dokkum}, P., {Brammer}, G.~B., {et~al.} 2019, \apj, 880,
  57, \dodoi{10.3847/1538-4357/ab290a}

\bibitem[{{Naab} {et~al.}(2009){Naab}, {Johansson}, \& {Ostriker}}]{Naab09}
{Naab}, T., {Johansson}, P.~H., \& {Ostriker}, J.~P. 2009, \apjl, 699, L178,
  \dodoi{10.1088/0004-637X/699/2/L178}

\bibitem[{{Neistein} {et~al.}(2006){Neistein}, {van den Bosch}, \&
  {Dekel}}]{Neistein06}
{Neistein}, E., {van den Bosch}, F.~C., \& {Dekel}, A. 2006, \mnras, 372, 933,
  \dodoi{10.1111/j.1365-2966.2006.10918.x}

\bibitem[{{Newman} {et~al.}(2012){Newman}, {Ellis}, {Bundy}, \&
  {Treu}}]{Newman12}
{Newman}, A.~B., {Ellis}, R.~S., {Bundy}, K., \& {Treu}, T. 2012, \apj, 746,
  162, \dodoi{10.1088/0004-637X/746/2/162}

\bibitem[{{Nipoti} {et~al.}(2009){Nipoti}, {Treu}, {Auger}, \&
  {Bolton}}]{Nipoti09}
{Nipoti}, C., {Treu}, T., {Auger}, M.~W., \& {Bolton}, A.~S. 2009, \apjl, 706,
  L86, \dodoi{10.1088/0004-637X/706/1/L86}

\bibitem[{{Nipoti} {et~al.}(2012){Nipoti}, {Treu}, {Leauthaud}, {Bundy},
  {Newman}, \& {Auger}}]{Nipoti12}
{Nipoti}, C., {Treu}, T., {Leauthaud}, A., {et~al.} 2012, \mnras, 422, 1714,
  \dodoi{10.1111/j.1365-2966.2012.20749.x}

\bibitem[{{Oh} {et~al.}(2017){Oh}, {Greene}, \& {Lackner}}]{Oh17}
{Oh}, S., {Greene}, J.~E., \& {Lackner}, C.~N. 2017, \apj, 836, 115,
  \dodoi{10.3847/1538-4357/836/1/115}

\bibitem[{{Oser} {et~al.}(2010){Oser}, {Ostriker}, {Naab}, {Johansson}, \&
  {Burkert}}]{Oser10}
{Oser}, L., {Ostriker}, J.~P., {Naab}, T., {Johansson}, P.~H., \& {Burkert}, A.
  2010, \apj, 725, 2312, \dodoi{10.1088/0004-637X/725/2/2312}

\bibitem[{{Ownsworth} {et~al.}(2014){Ownsworth}, {Conselice}, {Mortlock},
  {Hartley}, {Almaini}, {Duncan}, \& {Mundy}}]{Ownsworth14}
{Ownsworth}, J.~R., {Conselice}, C.~J., {Mortlock}, A., {et~al.} 2014, \mnras,
  445, 2198, \dodoi{10.1093/mnras/stu1802}

\bibitem[{{Paulino-Afonso} {et~al.}(2017){Paulino-Afonso}, {Sobral},
  {Buitrago}, \& {Afonso}}]{PaulinoAfonso17}
{Paulino-Afonso}, A., {Sobral}, D., {Buitrago}, F., \& {Afonso}, J. 2017,
  \mnras, 465, 2717, \dodoi{10.1093/mnras/stw2933}

\bibitem[{Pedregosa {et~al.}(2011)Pedregosa, Varoquaux, Gramfort, Michel,
  Thirion, Grisel, Blondel, Prettenhofer, Weiss, Dubourg,
  {et~al.}}]{pedregosa2011scikit}
Pedregosa, F., Varoquaux, G., Gramfort, A., {et~al.} 2011, Journal of machine
  learning research, 12, 2825

\bibitem[{{Planck Collaboration} {et~al.}(2016){Planck Collaboration}, {Ade},
  {Aghanim}, {Arnaud}, {Ashdown}, {Aumont}, {Baccigalupi}, {Banday},
  {Barreiro}, {Bartlett}, {Bartolo}, {Battaner}, {Battye}, {Benabed},
  {Beno{\^\i}t}, {Benoit-L{\'e}vy}, {Bernard}, {Bersanelli}, {Bielewicz},
  {Bock}, {Bonaldi}, {Bonavera}, {Bond}, {Borrill}, {Bouchet}, {Boulanger},
  {Bucher}, {Burigana}, {Butler}, {Calabrese}, {Cardoso}, {Catalano},
  {Challinor}, {Chamballu}, {Chary}, {Chiang}, {Chluba}, {Christensen},
  {Church}, {Clements}, {Colombi}, {Colombo}, {Combet}, {Coulais}, {Crill},
  {Curto}, {Cuttaia}, {Danese}, {Davies}, {Davis}, {de Bernardis}, {de Rosa},
  {de Zotti}, {Delabrouille}, {D{\'e}sert}, {Di Valentino}, {Dickinson},
  {Diego}, {Dolag}, {Dole}, {Donzelli}, {Dor{\'e}}, {Douspis}, {Ducout},
  {Dunkley}, {Dupac}, {Efstathiou}, {Elsner}, {En{\ss}lin}, {Eriksen},
  {Farhang}, {Fergusson}, {Finelli}, {Forni}, {Frailis}, {Fraisse},
  {Franceschi}, {Frejsel}, {Galeotta}, {Galli}, {Ganga}, {Gauthier}, {Gerbino},
  {Ghosh}, {Giard}, {Giraud-H{\'e}raud}, {Giusarma}, {Gjerl{\o}w},
  {Gonz{\'a}lez-Nuevo}, {G{\'o}rski}, {Gratton}, {Gregorio}, {Gruppuso},
  {Gudmundsson}, {Hamann}, {Hansen}, {Hanson}, {Harrison}, {Helou},
  {Henrot-Versill{\'e}}, {Hern{\'a}ndez-Monteagudo}, {Herranz}, {Hildebrandt},
  {Hivon}, {Hobson}, {Holmes}, {Hornstrup}, {Hovest}, {Huang}, {Huffenberger},
  {Hurier}, {Jaffe}, {Jaffe}, {Jones}, {Juvela}, {Keih{\"a}nen}, {Keskitalo},
  {Kisner}, {Kneissl}, {Knoche}, {Knox}, {Kunz}, {Kurki-Suonio}, {Lagache},
  {L{\"a}hteenm{\"a}ki}, {Lamarre}, {Lasenby}, {Lattanzi}, {Lawrence}, {Leahy},
  {Leonardi}, {Lesgourgues}, {Levrier}, {Lewis}, {Liguori}, {Lilje},
  {Linden-V{\o}rnle}, {L{\'o}pez-Caniego}, {Lubin}, {Mac{\'\i}as-P{\'e}rez},
  {Maggio}, {Maino}, {Mandolesi}, {Mangilli}, {Marchini}, {Maris}, {Martin},
  {Martinelli}, {Mart{\'\i}nez-Gonz{\'a}lez}, {Masi}, {Matarrese}, {McGehee},
  {Meinhold}, {Melchiorri}, {Melin}, {Mendes}, {Mennella}, {Migliaccio},
  {Millea}, {Mitra}, {Miville-Desch{\^e}nes}, {Moneti}, {Montier}, {Morgante},
  {Mortlock}, {Moss}, {Munshi}, {Murphy}, {Naselsky}, {Nati}, {Natoli},
  {Netterfield}, {N{\o}rgaard-Nielsen}, {Noviello}, {Novikov}, {Novikov},
  {Oxborrow}, {Paci}, {Pagano}, {Pajot}, {Paladini}, {Paoletti}, {Partridge},
  {Pasian}, {Patanchon}, {Pearson}, {Perdereau}, {Perotto}, {Perrotta},
  {Pettorino}, {Piacentini}, {Piat}, {Pierpaoli}, {Pietrobon}, {Plaszczynski},
  {Pointecouteau}, {Polenta}, {Popa}, {Pratt}, {Pr{\'e}zeau}, {Prunet},
  {Puget}, {Rachen}, {Reach}, {Rebolo}, {Reinecke}, {Remazeilles}, {Renault},
  {Renzi}, {Ristorcelli}, {Rocha}, {Rosset}, {Rossetti}, {Roudier},
  {Rouill{\'e} d'Orfeuil}, {Rowan-Robinson}, {Rubi{\~n}o-Mart{\'\i}n},
  {Rusholme}, {Said}, {Salvatelli}, {Salvati}, {Sandri}, {Santos},
  {Savelainen}, {Savini}, {Scott}, {Seiffert}, {Serra}, {Shellard}, {Spencer},
  {Spinelli}, {Stolyarov}, {Stompor}, {Sudiwala}, {Sunyaev}, {Sutton},
  {Suur-Uski}, {Sygnet}, {Tauber}, {Terenzi}, {Toffolatti}, {Tomasi},
  {Tristram}, {Trombetti}, {Tucci}, {Tuovinen}, {T{\"u}rler}, {Umana},
  {Valenziano}, {Valiviita}, {Van Tent}, {Vielva}, {Villa}, {Wade}, {Wandelt},
  {Wehus}, {White}, {White}, {Wilkinson}, {Yvon}, {Zacchei}, \&
  {Zonca}}]{Planck16}
{Planck Collaboration}, {Ade}, P.~A.~R., {Aghanim}, N., {et~al.} 2016, \aap,
  594, A13, \dodoi{10.1051/0004-6361/201525830}

\bibitem[{{Poggianti} \& {Barbaro}(1997)}]{Poggianti97}
{Poggianti}, B.~M., \& {Barbaro}, G. 1997, \aap, 325, 1025.
\newblock \doarXiv{astro-ph/9703067}

\bibitem[{{Saglia} {et~al.}(2010){Saglia}, {S{\'a}nchez-Bl{\'a}zquez},
  {Bender}, {Simard}, {Desai}, {Arag{\'o}n-Salamanca}, {Milvang-Jensen},
  {Halliday}, {Jablonka}, {Noll}, {Poggianti}, {Clowe}, {De Lucia},
  {Pell{\'o}}, {Rudnick}, {Valentinuzzi}, {White}, \& {Zaritsky}}]{Saglia10}
{Saglia}, R.~P., {S{\'a}nchez-Bl{\'a}zquez}, P., {Bender}, R., {et~al.} 2010,
  \aap, 524, A6, \dodoi{10.1051/0004-6361/201014703}

\bibitem[{{Scarlata} {et~al.}(2007){Scarlata}, {Carollo}, {Lilly}, {Sargent},
  {Feldmann}, {Kampczyk}, {Porciani}, {Koekemoer}, {Scoville}, {Kneib},
  {Leauthaud}, {Massey}, {Rhodes}, {Tasca}, {Capak}, {Maier}, {McCracken},
  {Mobasher}, {Renzini}, {Taniguchi}, {Thompson}, {Sheth}, {Ajiki}, {Aussel},
  {Murayama}, {Sanders}, {Sasaki}, {Shioya}, \& {Takahashi}}]{Scarlata07}
{Scarlata}, C., {Carollo}, C.~M., {Lilly}, S., {et~al.} 2007, \apjs, 172, 406,
  \dodoi{10.1086/516582}

\bibitem[{{Sersic}(1968)}]{Sersic68}
{Sersic}, J.~L. 1968, {Atlas de Galaxias Australes}

\bibitem[{{Shen} {et~al.}(2003){Shen}, {Mo}, {White}, {Blanton}, {Kauffmann},
  {Voges}, {Brinkmann}, \& {Csabai}}]{Shen03}
{Shen}, S., {Mo}, H.~J., {White}, S. D.~M., {et~al.} 2003, \mnras, 343, 978,
  \dodoi{10.1046/j.1365-8711.2003.06740.x}

\bibitem[{{Sohn} {et~al.}(2021){Sohn}, {Geller}, {Hwang}, {Fabricant}, {Moran},
  \& {Utsumi}}]{Sohn21}
{Sohn}, J., {Geller}, M.~J., {Hwang}, H.~S., {et~al.} 2021, \apj, 909, 129,
  \dodoi{10.3847/1538-4357/abd9be}

\bibitem[{{Sohn} {et~al.}(\noop{3001} 2022, in prep.){Sohn}, {Geller}, {Hwang},
  {Fabricant}, \& {Utsumi}}]{Sohn22}
{Sohn}, J., {Geller}, M.~J., {Hwang}, H.~S., {Fabricant}, D.~G., \& {Utsumi},
  Y. \noop{3001} 2022, in prep., \apj

\bibitem[{{Suess} {et~al.}(2019){Suess}, {Kriek}, {Price}, \&
  {Barro}}]{Suess19}
{Suess}, K.~A., {Kriek}, M., {Price}, S.~H., \& {Barro}, G. 2019, \apj, 877,
  103, \dodoi{10.3847/1538-4357/ab1bda}

\bibitem[{{Suess} {et~al.}(2022){Suess}, {Bezanson}, {Nelson}, {Setton},
  {Price}, {van Dokkum}, {Brammer}, {Labbe}, {Leja}, {Miller}, {Robertson},
  {Weaver}, \& {Whitaker}}]{Suess22}
{Suess}, K.~A., {Bezanson}, R., {Nelson}, E.~J., {et~al.} 2022, arXiv e-prints,
  arXiv:2207.10655.
\newblock \doarXiv{2207.10655}

\bibitem[{{Szomoru} {et~al.}(2013){Szomoru}, {Franx}, {van Dokkum}, {Trenti},
  {Illingworth}, {Labb{\'e}}, \& {Oesch}}]{Szomoru13}
{Szomoru}, D., {Franx}, M., {van Dokkum}, P.~G., {et~al.} 2013, \apj, 763, 73,
  \dodoi{10.1088/0004-637X/763/2/73}

\bibitem[{{Tacchella} {et~al.}(2017){Tacchella}, {Carollo}, {Faber}, {Cibinel},
  {Dekel}, {Koo}, {Renzini}, \& {Woo}}]{Tacchella17}
{Tacchella}, S., {Carollo}, C.~M., {Faber}, S.~M., {et~al.} 2017, \apjl, 844,
  L1, \dodoi{10.3847/2041-8213/aa7cfb}

\bibitem[{{Taylor} {et~al.}(2018){Taylor}, {Cirasuolo}, {Afonso}, {Carollo},
  {Evans}, {Flores}, {Maiolino}, {Paltani}, {Vanzi}, {Abreu}, {Amans},
  {Atkinson}, {Barrett}, {Beard}, {B{\'e}chet}, {Black}, {Boettger},
  {Brierley}, {Buscher}, {Cabral}, {Cochrane}, {Coelho}, {Colling},
  {Conzelmann}, {Dalessio}, {Dauvin}, {Davidson}, {Drass}, {D{\"u}nner},
  {Fairley}, {Fasola}, {Ferruzzi}, {Fisher}, {Flores}, {Garilli}, {Gaudemard},
  {Gonzalez}, {Guinouard}, {Gutierrez}, {Hammersley}, {Haigron}, {Haniff},
  {Hayati}, {Ives}, {Iwert}, {Laporte}, {Lee}, {Li Causi}, {Luco}, {Macleod},
  {Mainieri}, {Maire}, {Melse}, {Nix}, {Oliva}, {Oliveira}, {Origlia}, {Parry},
  {Pedichini}, {Piazzesi}, {Rees}, {Reix}, {Rodrigues}, {Rojas}, {Rota},
  {Royer}, {Santos}, {Schnell}, {Shen}, {Sordet}, {Strachan}, {Sun}, {Tait},
  {Torres}, {Tozzi}, {Tulloch}, {Navarro}, {Von Dran}, {Waring}, {Watson},
  {Woodward}, \& {Yang}}]{Taylor18}
{Taylor}, W., {Cirasuolo}, M., {Afonso}, J., {et~al.} 2018, in Society of
  Photo-Optical Instrumentation Engineers (SPIE) Conference Series, Vol. 10702,
  Ground-based and Airborne Instrumentation for Astronomy VII, ed. C.~J.
  {Evans}, L.~{Simard}, \& H.~{Takami}, 107021G, \dodoi{10.1117/12.2313403}

\bibitem[{{The Astropy Collaboration} {et~al.}(2022){The Astropy
  Collaboration}, {Price-Whelan}, {Lian Lim}, {Earl}, {Starkman}, {Bradley},
  {Shupe}, {Patil}, {Corrales}, {Brasseur}, {N{\"o}the}, {Donath}, {Tollerud},
  {Morris}, {Ginsburg}, {Vaher}, {Weaver}, {Tocknell}, {Jamieson}, {van
  Kerkwijk}, {Robitaille}, {Merry}, {Bachetti}, {G{\"u}nther}, {Aldcroft},
  {Alvarado-Montes}, {Archibald}, {B{\'o}di}, {Bapat}, {Barentsen},
  {Baz{\'a}n}, {Biswas}, {Boquien}, {Burke}, {Cara}, {Cara}, {E Conroy},
  {Conseil}, {Craig}, {Cross}, {Cruz}, {D'Eugenio}, {Dencheva}, {Devillepoix},
  {Dietrich}, {Davis Eigenbrot}, {Erben}, {Ferreira}, {Foreman-Mackey}, {Fox},
  {Freij}, {Garg}, {Geda}, {Glattly}, {Gondhalekar}, {Gordon}, {Grant},
  {Greenfield}, {Groener}, {Guest}, {Gurovich}, {Handberg}, {Hart},
  {Hatfield-Dodds}, {Homeier}, {Hosseinzadeh}, {Jenness}, {Jones}, {Joseph},
  {Bryce Kalmbach}, {Karamehmetoglu}, {Ka{\l}uszy{\'n}ski}, {Kelley}, {Kern},
  {Kerzendorf}, {Koch}, {Kulumani}, {Lee}, {Ly}, {Ma}, {MacBride}, {Maljaars},
  {Muna}, {Murphy}, {Norman}, {O'Steen}, {Oman}, {Pacifici}, {Pascual},
  {Pascual-Granado}, {Patil}, {Perren}, {Pickering}, {Rastogi}, {Roulston},
  {Ryan}, {Rykoff}, {Sabater}, {Sakurikar}, {Salgado}, {Sanghi}, {Saunders},
  {Savchenko}, {Schwardt}, {Seifert-Eckert}, {Shih}, {Shrey Jain}, {Shukla},
  {Sick}, {Simpson}, {Singanamalla}, {Singer}, {Singhal}, {Sinha},
  {Sip{\H{o}}cz}, {Spitler}, {Stansby}, {Streicher}, {{\v{S}}umak}, {Swinbank},
  {Taranu}, {Tewary}, {Tremblay}, {de Val-Borro}, {Van Kooten}, {Vasovi{\'c}},
  {Verma}, {Cardoso}, {Williams}, {Wilson}, {Winkel}, {Wood-Vasey}, {Xue},
  {Yoachim}, {ZHANG}, \& {Zonca}}]{Astropy22}
{The Astropy Collaboration}, {Price-Whelan}, A.~M., {Lian Lim}, P., {et~al.}
  2022, arXiv e-prints, arXiv:2206.14220.
\newblock \doarXiv{2206.14220}

\bibitem[{{Toomre}(1977)}]{Toomre77}
{Toomre}, A. 1977, in Evolution of Galaxies and Stellar Populations, ed. B.~M.
  {Tinsley} \& D.~C. {Larson}, Richard B.~Gehret, 401

\bibitem[{{Trujillo} {et~al.}(2007){Trujillo}, {Conselice}, {Bundy}, {Cooper},
  {Eisenhardt}, \& {Ellis}}]{Trujillo07}
{Trujillo}, I., {Conselice}, C.~J., {Bundy}, K., {et~al.} 2007, \mnras, 382,
  109, \dodoi{10.1111/j.1365-2966.2007.12388.x}

\bibitem[{{Trujillo} {et~al.}(2011){Trujillo}, {Ferreras}, \& {de La
  Rosa}}]{Trujillo11}
{Trujillo}, I., {Ferreras}, I., \& {de La Rosa}, I.~G. 2011, \mnras, 415, 3903,
  \dodoi{10.1111/j.1365-2966.2011.19017.x}

\bibitem[{{Trujillo} {et~al.}(2006){Trujillo}, {F{\"o}rster Schreiber},
  {Rudnick}, {Barden}, {Franx}, {Rix}, {Caldwell}, {McIntosh}, {Toft},
  {H{\"a}ussler}, {Zirm}, {van Dokkum}, {Labb{\'e}}, {Moorwood},
  {R{\"o}ttgering}, {van der Wel}, {van der Werf}, \& {van
  Starkenburg}}]{Trujillo06}
{Trujillo}, I., {F{\"o}rster Schreiber}, N.~M., {Rudnick}, G., {et~al.} 2006,
  \apj, 650, 18, \dodoi{10.1086/506464}

\bibitem[{{van der Wel} {et~al.}(2014){van der Wel}, {Franx}, {van Dokkum},
  {Skelton}, {Momcheva}, {Whitaker}, {Brammer}, {Bell}, {Rix}, {Wuyts},
  {Ferguson}, {Holden}, {Barro}, {Koekemoer}, {Chang}, {McGrath},
  {H{\"a}ussler}, {Dekel}, {Behroozi}, {Fumagalli}, {Leja}, {Lundgren},
  {Maseda}, {Nelson}, {Wake}, {Patel}, {Labb{\'e}}, {Faber}, {Grogin}, \&
  {Kocevski}}]{vanderWel14}
{van der Wel}, A., {Franx}, M., {van Dokkum}, P.~G., {et~al.} 2014, \apj, 788,
  28, \dodoi{10.1088/0004-637X/788/1/28}

\bibitem[{{van Dokkum} \& {Franx}(2001)}]{vanDokkum01}
{van Dokkum}, P.~G., \& {Franx}, M. 2001, \apj, 553, 90, \dodoi{10.1086/320645}

\bibitem[{{van Dokkum} {et~al.}(2010){van Dokkum}, {Whitaker}, {Brammer},
  {Franx}, {Kriek}, {Labb{\'e}}, {Marchesini}, {Quadri}, {Bezanson},
  {Illingworth}, {Muzzin}, {Rudnick}, {Tal}, \& {Wake}}]{vanDokkum10}
{van Dokkum}, P.~G., {Whitaker}, K.~E., {Brammer}, G., {et~al.} 2010, \apj,
  709, 1018, \dodoi{10.1088/0004-637X/709/2/1018}

\bibitem[{{Vergani} {et~al.}(2008){Vergani}, {Scodeggio}, {Pozzetti}, {Iovino},
  {Franzetti}, {Garilli}, {Zamorani}, {Maccagni}, {Lamareille}, {Le F{\`e}vre},
  {Charlot}, {Contini}, {Guzzo}, {Bottini}, {Le Brun}, {Picat}, {Scaramella},
  {Tresse}, {Vettolani}, {Zanichelli}, {Adami}, {Arnouts}, {Bardelli},
  {Bolzonella}, {Cappi}, {Ciliegi}, {Foucaud}, {Gavignaud}, {Ilbert},
  {McCracken}, {Marano}, {Marinoni}, {Mazure}, {Meneux}, {Merighi}, {Paltani},
  {Pell{\`o}}, {Pollo}, {Radovich}, {Zucca}, {Bondi}, {Bongiorno},
  {Brinchmann}, {Cucciati}, {de la Torre}, {Gregorini}, {Perez-Montero},
  {Mellier}, {Merluzzi}, \& {Temporin}}]{Vergani08}
{Vergani}, D., {Scodeggio}, M., {Pozzetti}, L., {et~al.} 2008, \aap, 487, 89,
  \dodoi{10.1051/0004-6361:20077910}

\bibitem[{Virtanen {et~al.}(2020)Virtanen, Gommers, Oliphant, Haberland, Reddy,
  Cournapeau, Burovski, Peterson, Weckesser, Bright, {van der Walt}, Brett,
  Wilson, Millman, Mayorov, Nelson, Jones, Kern, Larson, Carey, Polat, Feng,
  Moore, {VanderPlas}, Laxalde, Perktold, Cimrman, Henriksen, Quintero, Harris,
  Archibald, Ribeiro, Pedregosa, {van Mulbregt}, \& {SciPy 1.0
  Contributors}}]{2020SciPy-NMeth}
Virtanen, P., Gommers, R., Oliphant, T.~E., {et~al.} 2020, Nature Methods, 17,
  261, \dodoi{10.1038/s41592-019-0686-2}

\bibitem[{{White}(1978)}]{White78}
{White}, S.~D.~M. 1978, \mnras, 184, 185, \dodoi{10.1093/mnras/184.2.185}

\bibitem[{{Whitney} {et~al.}(2021){Whitney}, {Ferreira}, {Conselice}, \&
  {Duncan}}]{Whitney21}
{Whitney}, A., {Ferreira}, L., {Conselice}, C.~J., \& {Duncan}, K. 2021, \apj,
  919, 139, \dodoi{10.3847/1538-4357/ac1422}

\bibitem[{{Williams} {et~al.}(2010){Williams}, {Quadri}, {Franx}, {van Dokkum},
  {Toft}, {Kriek}, \& {Labb{\'e}}}]{Williams10}
{Williams}, R.~J., {Quadri}, R.~F., {Franx}, M., {et~al.} 2010, \apj, 713, 738,
  \dodoi{10.1088/0004-637X/713/2/738}

\bibitem[{{Wu} {et~al.}(2018){Wu}, {van der Wel}, {Bezanson}, {Gallazzi},
  {Pacifici}, {Straatman}, {Bari{\v{s}}i{\'c}}, {Bell}, {Chauke}, {van Houdt},
  {Franx}, {Muzzin}, {Sobral}, \& {Wild}}]{Wu18}
{Wu}, P.-F., {van der Wel}, A., {Bezanson}, R., {et~al.} 2018, \apj, 868, 37,
  \dodoi{10.3847/1538-4357/aae822}

\bibitem[{{Yang} {et~al.}(2021){Yang}, {Roberts-Borsani}, {Treu}, {Birrer},
  {Morishita}, \& {Brada{\v{c}}}}]{Yang21}
{Yang}, L., {Roberts-Borsani}, G., {Treu}, T., {et~al.} 2021, \mnras, 501,
  1028, \dodoi{10.1093/mnras/staa3713}

\bibitem[{{Zahid} {et~al.}(2016){Zahid}, {Damjanov}, {Geller}, {Hwang}, \&
  {Fabricant}}]{Zahid16}
{Zahid}, H.~J., {Damjanov}, I., {Geller}, M.~J., {Hwang}, H.~S., \&
  {Fabricant}, D.~G. 2016, \apj, 821, 101, \dodoi{10.3847/0004-637X/821/2/101}

\bibitem[{{Zahid} \& {Geller}(2017)}]{Zahid17}
{Zahid}, H.~J., \& {Geller}, M.~J. 2017, \apj, 841, 32,
  \dodoi{10.3847/1538-4357/aa7056}

\bibitem[{{Zahid} {et~al.}(2019){Zahid}, {Geller}, {Damjanov}, \&
  {Sohn}}]{Zahid19}
{Zahid}, H.~J., {Geller}, M.~J., {Damjanov}, I., \& {Sohn}, J. 2019, \apj, 878,
  158, \dodoi{10.3847/1538-4357/ab21b9}

\bibitem[{{Zhou} {et~al.}(2022){Zhou}, {Dey}, {Newman}, {Eisenstein}, {Dawson},
  {Bailey}, {Berti}, {Guy}, {Lan}, {Zou}, {Aguilar}, {Ahlen}, {Alam}, {Brooks},
  {de la Macorra}, {Dey}, {Dhungana}, {Fanning}, {Font-Ribera}, {Gontcho},
  {Honscheid}, {Ishak}, {Kisner}, {Kov{\'a}cs}, {Kremin}, {Landriau}, {Levi},
  {Magneville}, {Martini}, {Meisner}, {Miquel}, {Moustakas}, {Myers}, {Nie},
  {Palanque-Delabrouille}, {Percival}, {Poppett}, {Prada}, {Raichoor}, {Ross},
  {Schlafly}, {Schlegel}, {Schubnell}, {Tarl{\'e}}, {Weaver}, {Wechsler},
  {Y{\`e}che}, \& {Zhou}}]{Zhou22}
{Zhou}, R., {Dey}, B., {Newman}, J.~A., {et~al.} 2022, arXiv e-prints,
  arXiv:2208.08515.
\newblock \doarXiv{2208.08515}

\end{thebibliography}
 \newcommand{\noop}[1]{}

\end{document}